\pdfoutput=1
\RequirePackage{ifpdf}
\ifpdf % We~are running pdfTeX in pdf mode
\documentclass[pdftex]{sigma}
\else
\documentclass{sigma}
\fi

\numberwithin{equation}{section}

\newtheorem{Theorem}{Theorem}[section]
\newtheorem*{Theorem*}{Theorem}

\theoremstyle{definition}

\newtheorem{Example}[Theorem]{Example}

\newcommand{\CA}{{\cal A}}
\newcommand{\CB}{{\cal B}}
\newcommand{\CC}{{\cal C}}

\newcommand{\CF}{{\cal F}}
\newcommand{\CG}{{\cal G}}
\newcommand{\CH}{{\cal H}}

\newcommand{\CK}{{\cal K}}
\newcommand{\CL}{{\cal L}}
\newcommand{\CM}{{\cal M}}

\newcommand{\CO}{{\cal O}}

\newcommand{\CR}{{\cal R}}
\newcommand{\CS}{{\cal S}}
\newcommand{\CT}{{\cal T}}

\def\IZ{{\mathbb Z}}
\def\IR{{\mathbb R}}

\def\IP{{\mathbb P}}

\newcommand{\re}{{\rm e}}
\newcommand{\ri}{{\rm i}}
\newcommand{\rd}{{\rm d}}

\newcommand{\mO}{\mathsf{O}}
\newcommand{\mR}{\mathsf{R}}

\newcommand{\mD}{\mathfrak{D}}
\newcommand{\oD}{\mathsf{D}}
\newcommand{\mW}{\mathsf{W}}
\newcommand{\be}{\begin{equation}}
\newcommand{\ee}{\end{equation}}
\newcommand{\ba}{\begin{aligned}}
\newcommand{\ea}{\end{aligned}}
\newcommand{\ben}{\begin{eqnarray}\displaystyle}
\newcommand{\een}{\end{eqnarray}}

\newdimen\tableauside\tableauside=1.0ex
\newdimen\tableaurule\tableaurule=0.4pt
\newdimen\tableaustep
\def\phantomhrule#1{\hbox{\vbox to0pt{\hrule height\tableaurule width#1\vss}}}
\def\phantomvrule#1{\vbox{\hbox to0pt{\vrule width\tableaurule height#1\hss}}}
\def\sqr{\vbox{%
 \phantomhrule\tableaustep
 \hbox{\phantomvrule\tableaustep\kern\tableaustep\phantomvrule\tableaustep}%
 \hbox{\vbox{\phantomhrule\tableauside}\kern-\tableaurule}}}
\def\squares#1{\hbox{\count0=#1\noindent\loop\sqr
 \advance\count0 by-1 \ifnum\count0>0\repeat}}
\def\tableau#1{\vcenter{\offinterlineskip
 \tableaustep=\tableauside\advance\tableaustep by-\tableaurule
 \kern\normallineskip\hbox
 {\kern\normallineskip\vbox
 {\gettableau#1 0 }%
 \kern\normallineskip\kern\tableaurule}%
 \kern\normallineskip\kern\tableaurule}}
\def\gettableau#1{\ifnum#1=0\let\next=\null\else
\squares{#1}\let\next=\gettableau\fi\next}

\newcommand{\bea}{\begin{eqnarray}\displaystyle}
\newcommand{\eea}{\end{eqnarray}}

\begin{document}

\allowdisplaybreaks

\newcommand{\arXivNumber}{2309.12046}

\renewcommand{\PaperNumber}{039}

\FirstPageHeading

\ShortArticleName{Non-Perturbative Real Topological Strings}

\ArticleName{Non-Perturbative Real Topological Strings}

\Author{Marcos MARI\~NO and Maximilian SCHWICK}

\AuthorNameForHeading{M.~Mari\~no and M.~Schwick}

\Address{D\'epartement de Physique Th\'eorique et Section de Math\'ematiques,\\
Universit\'e de Gen\`eve, Gen\`eve, 1211, Switzerland}
\Email{\mail{marcos.marino@unige.ch}, \mail{maximilian.schwick@unige.ch}}
\URLaddress{\url{http://www.marcosmarino.net}}

\ArticleDates{Received September 11, 2025, in final form March 23, 2026; Published online April 20, 2026}

\Abstract{We study the resurgent structure of Walcher's real topological string on general Calabi--Yau manifolds. We find trans-series solutions to the corresponding holomorphic anomaly equations, at all orders in the string coupling constant, by extending the operator formalism of the closed topological string, and we obtain explicit formulae for multi-instanton amplitudes. We find that the integer invariants counting disks appear as Stokes constants in the resurgent structure, and we provide experimental evidence for our results in the case of the real topological string on local $\mathbb{P}^2$.}


\Classification{14N35; 34M40}

\section{Introduction}

Topological string theory is defined perturbatively, and it has been a~longstanding problem to understand
its non-perturbative structure. An important clue for this is
the asymptotic behavior of its genus expansion: as in
other string theories~\cite{gp,shenker}, it grows factorially (see, e.g.,~\cite{mmlargen}).
This usually indicates the existence of exponentially small corrections
in the string coupling constant. In the case of
conventional strings and minimal strings, it is believed that such effects
are due to D-branes~\cite{polchinski}.

In~\cite{mmnp, mmopen,msw}, it was proposed that a systematic understanding
of the non-perturbative effects associated
to the divergence of the topological string perturbation series can be obtained
by using the theory of resurgence
(see~\cite{abs, mmlargen, mmbook,msauzin, ss} for introductions and reviews). According to
this theory, one can associate to the perturbative series a collection of exponentially
small corrections, describing non-perturbative amplitudes,
together with numerical invariants called Stokes constants (see, e.g.,~\cite{gm-peacock}
for a definition of such a ``resurgent
structure''). In addition, there is growing evidence that the Stokes constants characterizing
the resurgent structure of the topological string
free energy are closely related to BPS invariants of the Calabi--Yau (CY) threefold
\cite{astt, ghn, gu-res, gkkm, gm-peacock, gm-multi, im, rella} (a connection between
Stokes constants and BPS invariants has been also found in complex
Chern--Simons theory and its supersymmetric 3d duals, see~\mbox{\cite{ggm1,ggm2,ggmw,gmp,wheeler}}).

The resurgent structure of topological string theory is therefore a very rich object,
containing perhaps information about all the BPS
invariants of the theory, and a complete description is still lacking. It is possible
however to obtain explicit expressions for the non-perturbative corrections by
using trans-series solutions to the holomorphic anomaly equations (HAEs) of
\cite{bcov-pre,bcov}. This idea was put forward in~\cite{cesv2, cesv1} in the local
CY case. It was further developed in~\cite{gm-multi}, where exact trans-series solutions
in closed form were obtained, and extended to compact CY manifolds in~\cite{gkkm}.
The results of~\cite{gkkm, gm-multi} are based on an operator formalism first considered
in~\cite{codesido-thesis, coms}. One finds in particular that multi-instanton solutions to the HAE are given by a~generalization
of the eigenvalue tunneling of matrix models~\cite{david, msw,multi-multi,shenker},
which suggests that flat coordinates are quantized in integer units of the string coupling constant.

All the studies done so far focused on closed topological strings. There is a very interesting and rich generalization
thereof which we will call the real topological string. Its
theory was developed in a series of papers~\cite{kw-localp2, walcher-dq, walcher-tadpole, walcher-bcov}
by J. Walcher and collaborators. The basic idea is to introduce a real three-dimensional locus
in the CY target which plays two simultaneous roles: on the one hand it is a submanifold wrapped
by a D-brane, giving boundary conditions for open topological strings,
and on the other hand it is an orientifold plane, leading in addition to non-orientable amplitudes.
Originally, the theory was conceived as involving
open and closed strings only~\cite{walcher-dq, walcher-bcov}, but it was later realized that consistency
requires the non-orientable sector as well~\cite{walcher-tadpole}.

The real topological string leads to a perturbative series in the string coupling
constant due to a sum over orientable and non-orientable
surfaces, with and without boundaries. It~is then natural to ask what is the resurgent
structure of this series and whether the
results of~\cite{gkkm, gm-multi} can be generalized to this setting. In this paper, we start a systematic
exploration of these issues. First, we generalize the results of~\cite{gkkm, gm-multi} and we obtain
trans-series solutions to Walcher's HAE~\cite{walcher-tadpole, walcher-bcov}
for the real topological string. This is achieved by extending the operator formalism
of~\cite{gkkm, gm-multi}. The trans-series can be written down for arbitrary holomorphic instanton actions. As in
previous studies~\cite{cesv2, cesv1, dmp-np, gkkm, gm-multi}, we find that the relevant actions are, in the
B-model language, integral periods of the
holomorphic 3-form (in the A-model they can be interpreted as central charges of D-branes.)
In addition to these, we find that half-integral periods play also a r\^ole, presumably due to
the orientifold action. One consequence of our analysis is that multi-instanton solutions
to the HAE of the real topological string can be again interpreted in terms of shifts of the closed string background
by integer multiples of the string coupling constant. The boundary conditions for the trans-series are provided
by the behavior of the real topological string amplitudes at the conifold and the large radius points.
As usual in resurgence theory, our results make it possible to obtain the asymptotics of the free energies
for large ``genus'' or Euler characteristic of the worldsheet, which we check successfully against perturbative results
in the case of the real topological string on local $\IP^2$. In addition, we find indications that
the connection between Stokes constants and BPS invariants extends to the
real topological string, and one can explicitly show that the integer invariants counting disks appear as Stokes constants.

This paper is organized as follows. In Section~\ref{review-TS}, we review relevant results about the real topological string, focusing
on Walcher's HAE. Some of the results on the real propagators, like, e.g., (\ref{gamma-real}), seem however to be new.
In Section~\ref{sec:trans-series-res-structure}, we study
the non-perturbative aspects of the theory. We construct trans-series solutions to the HAE by extending the operator formalism
of~\cite{gkkm, gm-multi} to the real case, and we study different boundary conditions for these equations. In~Section~\ref{sec-p2}, we test the one-instanton amplitude obtained in Section~\ref{sec:trans-series-res-structure} in the
case of local $\IP^2$. Finally, Section~\ref{sec-conclusions}
contains conclusions and prospects for future studies. Appendix \ref{app-localp2} collects useful results
for local $\IP^2$ which are used in the paper.

\section{The real topological string}
\label{review-TS}
\subsection{Free energy and holomorphic anomaly equations}

The real topological string involves a CY manifold $X$ together with a D-brane configuration and an orientifold plane. In practice,
we consider an antiholomorphic involution of $X$,
\be
\sigma\colon\ X \rightarrow X.
\ee
The fixed locus $L$ of $\sigma$ is a Lagrangian submanifold. It can be wrapped by an A-brane, after a~choice
of an $U(1)$ flat bundle on it, but we can also consider $L$ as an orientifold plane. In the real
topological string, we have to include the A-brane and the orientifold plane at the same time,
as explained in~\cite{walcher-tadpole}.

\begin{Example} The main example in this paper will be the real topological string on local $\IP^2$, first considered
in~\cite{walcher-tadpole}. In this
case, $X$ is the total space of the bundle $\CO(-3) \rightarrow \IP^2$. The involution
$\sigma$ acts by
complex conjugation on both the fiber and the base. The fixed locus $L$ is the total space of the
orientation bundle over $\IR\IP^2$, and since it has $H_1(L, \IZ) =\IZ_2$, there are two choices for a
flat $U(1)$ bundle (or Wilson line). Another example is the quintic CY~\cite{cdgp}, together with an
involution which acts again by complex conjugation. Its fixed point locus $L$ is topologically a real
projective space $\IR \IP^3$, and $H_1(L, \IZ) =\IZ_2$ as well (see, e.g.,~\cite{walcher-dq}).
\end{Example}

In the real topological string, one has to consider contributions from all orientable and non-orientable
worldsheet topologies, with and without boundaries. The topological class of the worldsheet is determined by the
genus $g\ge 0$, the number of
crosscaps $c=0,1,2$, and the number of boundaries $h\ge 0$. The Euler characteristic of such a surface, $\chi$, is given by
\be
\chi=2g-2+h+c.
\ee
The total free energy is obtained as a sum over all possible values of $\chi \ge -2$, and
it has the form
\begin{gather}
\label{eq:perturbativeRealFreeEnergy}
\CG= \sum_{\chi \ge -2} \CG_{\chi} g_s^{\chi}.
\end{gather}
We will denote by $ \CF_{g,h}$ the contribution to $\CG_\chi$ coming from
orientable Riemann surfaces with genus $g$ and $h$ boundaries, and by \smash{$\CR_{g,h}$},
\smash{$\CK_{g,h}$} the contribution from non-orientable Riemann surfaces of genus $g$ with $h$ boundaries,
and $c=1,2$, respectively.\footnote{Our labelling of non-orientable
Riemann surfaces is different from the one used in, e.g.,~\cite{kw-localp2}.} We then
have
\be
\CG=\sum_{\chi=2g-2+h} \CF_{g,h} g_s^{\chi}+ \sum_{\chi=2g-1+h}\CR_{g,h}g_s^{\chi}+ \sum_{\chi=2g+h}\CK_{g,h}g_s^{\chi}.
\ee

In the examples studied so far~\cite{kw-localp2, walcher-tadpole}, and the ones considered in this paper,
one has that $\CR_{g,h}=0$ due to the tadpole cancellation condition of~\cite{walcher-tadpole}.\footnote{These amplitudes might be however
nonzero for more complicated situations, but this is well beyond the scope of this paper. We thank J.~Walcher for clarifications on this point.}

The real topological string includes in particular contributions from the conventional closed
topological string, corresponding to $c=h=0$. The first new ingredient is the disk amplitude with $g=c=0$, $h=1$,
which we will sometimes denote as
\be
\CG_{-1} = \CT.
\ee
When $H_1(L, \IZ) =\IZ_2$, as in the examples we will consider, we can interpret $\CT$
as the tension of a BPS domain wall interpolating between the two vacua on the D-brane~\cite{walcher-dq}.

In the A-model, the real
topological string provides a generalization of Gromov--Witten theory which ``counts''
holomorphic maps from open and non-orientable Riemann surfaces to the target space $X$. We recall that, in conventional
closed topological string theory, the enumerative information of the Gromov--Witten invariants
can be repackaged in terms of {\it integer} BPS invariants~\cite{gv}, and the total closed string free energy
\be
\CF(t;g_s)=\sum_{g \ge 0} \CF_{g} (t)g_s^{2g-2}
\ee
can be written as
\be
\label{gv-F}
\CF(t;g_s) = \sum_{d \in H_2(X,\IZ)}\sum_{r \ge 0} \sum_{k=1}^\infty n_{r,d} {1\over k}\biggl( 2 \sin {k g_s \over 2} \biggr)^{2r-2} \re^{-k d \cdot t},
\ee
up to a cubic polynomial in the $t_i$s. Here $t=(t_1, \dots, t_s)$ is the vector of
complexified K\"ahler parameters of $X$, and $n_{r,d}$ are the
Gopakumar--Vafa invariants, which are integers (see~\cite{dw} for a
detailed derivation of (\ref{gv-F}) from a physics perspective).
The integrality structure in (\ref{gv-F}) can be extended to the real topological string, and it
was proposed in~\cite{kw,walcher-tadpole,walcher-bcov} that
\be
\label{BPS-real}
\CG- \CF = 2\sum_{r \ge -1} \sum_{d \in H_2(X,\IZ) \atop d_i \equiv r \, \text{mod}\, 2}
\sum_{k \, \text{odd}} \tilde n_{r,d} {1\over k} \biggl( 2 \ri \sin {k g_s \over 2} \biggr)^r \re^{-d k \cdot t/2}.
\ee
Here, $\tilde n_{r,d}$ are new integer invariants counting BPS states. An M-theoretic derivation of this
integrality formula has been proposed in~\cite{pu}. As it was
emphasized in~\cite{walcher-tadpole}, one has
to combine the open and the non-orientable sector to obtain integer invariants $\tilde n_{r, d}$. Note
however that, when $r=-1$, only disks contribute, and $\tilde n_{-1,d}$ can be regarded as an integer counting of
disks in $X$ with boundary conditions set by $L$.

As one would expect from mirror symmetry, the calculation of $\CG_\chi$ is easier in the B-model.
This was first shown in the case of $\CT=\CG_{-1}$ in~\cite{walcher-dq}. The domain wall tension
is a holomorphic section of the Hodge line bundle $\CL$
over the moduli space of complex structures $\CM$ of $X$. Let us denote
by $z^a$, with $a=1,\dots, s$, a set of generic complex coordinates on this moduli space, where $s$ is the dimension of $\CM$.
Then, $\CT$ solves an inhomogeneous Picard--Fuchs equation of the form
\be
\label{inh-pf}
\mathfrak{L} \CT= f(z^a),
\ee
where $\mathfrak{L}$ is the Picard--Fuchs operator which governs the closed string periods,
and $f(z^a)$ is a~known function of the
complex moduli. This leads to an efficient counting of disks ending on~$L$.
The question then arises of how to compute the higher order terms in the real topological
string free energy. This problem
was solved by Walcher by finding {\it extended} HAEs for the real topological string, which generalize
the closed string case originally studied in~\cite{bcov-pre,bcov}. In order to present Walcher's HAE we need to
recall some basic ingredients on the special geometry of the
CY moduli space $\CM$ (further details can be found in, e.g.,~\cite{bcov, cdlo,klemm}).

The moduli space $\CM$ of complex structures of the mirror CY is a special
K\"ahler manifold of complex dimension $s$. The K\"ahler potential will be denoted by $K$, and
\be
G_{a \bar b} =\partial_a \partial_{\bar b} K
\ee
is the corresponding K\"ahler metric. The line bundle
$\CL$ over $\CM$ is endowed with a $U(1)$ connection $A_a= \partial_a K\equiv K_a$, and the covariant
derivative $D_a$ involves both the Levi-Civita connection associated to the metric,
\be
\Gamma^a_{bc} = G^{a \bar k} \partial_b G_{c \bar k},
\ee
and the $U(1)$ connection on $\CL$. Since $\CM$ is a special K\"ahler manifold, its
Christoffel symbols satisfy the special geometry relation
\be
 \label{anti-chris}
 \partial_{\bar b} \Gamma_{ac}^d= G_{a\bar b} \delta_c^d + G_{ c \bar b} \delta_a^d - C_{acm} \overline C_{\bar b}^{md},
 \ee
where $C_{abc}$ is the Yukawa coupling, and
\be
\label{antic}
{\overline C}_{\bar k}^{ij}= \re^{2 K} \, {\overline C}_{\bar k \bar a \bar b} G^{i \bar a} G^{j \bar b}.
\ee
We recall that, if $\Omega$ is the holomorphic three-form of $X$, the Yukawa coupling is defined by
\be
\label{yukawa}
C_{abc}=\int_X \Omega \wedge {\partial^3 \Omega \over \partial z^a \partial z^b \partial z^c}.
\ee
In topological string theory, it corresponds to the two-sphere amplitude with three insertions.

In the B-model, the free energies $\CG_\chi$ get promoted in general
to non-holomorphic quantities which we will denote by $G_\chi$ (as in~\cite{gm-multi}, non-holomorphic and holomorphic versions of the
free energies will written as capital Roman and capital calligraphic letters, respectively).
The $G_\chi$ are sections of the bundle $\CL^{-\chi}$,
and they satisfy Walcher's HAE. The first ingredient we will need to set up these
HAE is the disk two-point function $\Delta_{ij}$, where $i$, $j$ are moduli indices. Morally, we have
\be
\Delta_{ij} \sim D_i D_j G_{-1}.
\ee
A more detailed analysis shows that $\Delta_{ij}$ is given by the so-called Griffiths'
infinitesimal invariant~\cite{walcher-bcov}. It can be written in terms of the domain wall tension $\CT$ as
\be
\label{griff}
\Delta_{ij}=D_i D_j \CT-\ri C_{ijk} \re^K G^{k \bar m} D_{\bar m} \overline{\CT}.
\ee
The Griffiths invariant is not holomorphic, and by using (\ref{anti-chris}) one finds that it satisfies the~HAE%
\be
\label{del-del}
\partial_{\bar a} \Delta_{ij} = -\ri C_{ijp} \re^K G^{p \bar b} \Delta_{\bar a \bar b},
\ee
where
\be
\Delta_{\bar a \bar b}= \overline{\Delta_{a b}}.
\ee
The Griffiths' invariant plays the r\^ole in the open string sector of the Yukawa coupling in the closed string sector.
For this reason, it is convenient to introduce its anti-holomorphic version~as%
\be
\Delta_{\bar a}^{~ b}= \ri \re^K G^{b \bar c} \Delta_{\bar a \bar c}.
\ee

We are now ready to consider the HAE for the real topological string in the B-model. As in the closed string case, the one-loop
contribution corresponding to $\chi=0$ is somewhat special and needs a separate treatment. There are three different contributions
at $\chi=0$: the torus amplitude $F_1$, the annulus amplitude $F_{0,2}$, and the
Klein bottle amplitude $K_{0,0}$.\footnote{Non-holomorphic Klein bottle amplitudes $K_{g,h}$ shouldn't be confused with the
K\"ahler potential $K$ and its derivatives $K_i$, $K_{ij}$.} We first
recall that the closed string amplitude at genus one satisfies the HAE~\cite{bcov-pre},
\begin{equation}
\partial_{\bar k} \partial_m F_1={1\over 2}
{\overline C}_{\bar k}^{ij} C_{mij} - \Bigl( {\chi\over 24} -1\Bigr)
G_{\bar km},
\label{holan1}\
\end{equation}
where $\chi$ is the Euler number of the CY $X$. The annulus amplitude satisfies~\cite{walcher-tadpole, walcher-bcov}
\be
\label{annulus-HAE}
\partial_{\bar a} \partial_j F_{0,2}= - \Delta_{jk} \Delta_{\bar a}^{~k}.
\ee
Finally, the Klein bottle amplitude satisfies, for the type of geometries we will consider~\cite{walcher-tadpole},
\be
 \partial_{\bar k} \partial_m K_{0,0} = {1\over 2}
{\overline C}_{\bar k}^{ij} C_{mij}-G_{\bar k m},
\ee
which is very similar to the HAE for $F_1$. In the next section, we will solve these equations explicitly in terms of propagators.

We can now write down Walcher's extended
HAE for the real topological string free energies with $\chi \ge 1$~\cite{walcher-tadpole, walcher-bcov}. It is given by
\be
\label{real-hae}
\partial_{\bar a} G_{\chi}= {1\over 2} \overline C_{\bar a}^{P\, jk} \sum\limits_{{\chi_1+\chi_2=\chi-2}\atop{\chi_1,\chi_2\geq 0}} D_j G_{\chi_1} D_k G_{\chi_2} + \overline C_{\bar a}^{P\, jk} D_j D_k G_{\chi-2}- \Delta_{\bar a}^{P\, j} D_j G_{\chi-1},
\ee
and it applies for $\chi \ge 1$. In these equations, \smash{$\overline C_{\bar a}^{P\, jk} $} and \smash{$\Delta_{\bar a}^{P~ j}$} are defined as~\cite{walcher-tadpole}
\be
\overline C_{\bar a}^{P\, jk} = \overline C_{\bar a}^{jl} {\delta_l^k +P_l^k \over 2}, \qquad
\Delta_{\bar a}^{P\, k}=\Delta_{\bar a}^{ j}{\delta_j^k +P_j^k \over 2},
\ee
and $P_l^k$ is a projector related to the orientifold action. In the cases we will consider,
$P_l^k= \delta_l^k$, and \smash{$\overline C_{\bar a}^{P\, jk} =
\overline C_{\bar a}^{jk}$, $\Delta_{\bar a}^{P\, j}=\Delta_{\bar a}^{ j}$}.
As in the case of the
closed topological string, the HAE can~be~solved recursively starting from the free energy with $\chi=0$ and the ``tree level data'' given by the
Yukawa coupling and Griffith's invariant. The best
procedure to solve these equations is probably the so-called ``direct integration'' method, developed, e.g., in~\cite{al,gkmw,hk06, yy}.
To develop this method, one has to use propagators, which will also play a crucial role in understanding the non-perturbative sectors.
They are the subject of the next section.

\subsection{Propagators }

Propagators were introduced in~\cite{bcov} as a tool to solve the HAE for the closed topological string, and they were
generalized in~\cite{walcher-bcov} to solve for (\ref{real-hae}). In this section we will provide a detailed description of the
propagators in both the real and the closed case (for the closed case one might consult~\cite{gkkm, hosono}).
Some of the results for the
real propagators seem to be new, and they will be needed in the discussion of the
trans-series solution to the extended HAE.

The two-index, closed string propagator is defined by~\cite{bcov}
\be
\label{propa-def}
\partial_{\bar k} S^{ij}=\re^{2 K} \, G^{i \bar a} G^{j \bar b} {\overline C}_{\bar k \bar a \bar b} .
\ee
 In addition to $S^{ab}$, one introduces as well
\be
\label{propa2-def}
\partial_{\bar c} S^b = G_{a \bar c} S^{ab}, \qquad
\partial_{\bar c } S = G_{a \bar c} S^a.
\ee
As in~\cite{gkkm}, we will use the propagator formalism set up in~\cite{al}. We introduce the shifted propagators
\begin{align}
\begin{aligned}
&\tilde{S}^{ij}= S^{ij}, \\
&\tilde{S}^i= S^i - S^{ij} K_j, \\
&\tilde{S}= S- S^i K_i + \frac{1}{2} S^{ij} K_i K_j.
\end{aligned}\label{shift}
\end{align}
A fundamental result of~\cite{al, bcov,gkmw} is that the derivatives of the propagators with respect to the moduli can be
written as quadratic expressions in the propagators, and one has the equations
\begin{align}
\begin{aligned}
&\partial_i S^{jk}= C_{imn} S^{mj} S^{nk} + \delta_i^j \tilde S^k +\delta_i^k \tilde S^j-s_{im}^j S^{mk} -s_{im}^k S^{mj} + h_i^{jk}, \\
&\partial_i \tilde S^j= C_{imn} S^{mj} \tilde S^n + 2 \delta_i^j \tilde S -s_{im}^j \tilde S^m -h_{ik} S^{kj} +h_i^j, \\
&\partial_i \tilde S= \frac{1}{2} C_{imn} \tilde S^m \tilde S^n -h_{ij} \tilde S^j +h_i, \\
&\partial_i K_j= K_i K_j -C_{ijn}S^{mn} K_m + s_{ij}^m K_m -C_{ijk} \tilde S^k + h_{ij}.
\end{aligned}\label{ders-tilded}
\end{align}
In these equation, $s_{ac}^r$, $h_i^{jk}$, $h_{ij}$, $h_i^j$ and $h_i$ are
holomorphic functions or ambiguities, and they are usually chosen to be rational functions of the moduli.
We also have the following important relation between the Christoffel symbols of
the K\"ahler metric and the propagator $S^{ab}$:
\be
\label{chris-prop}
\Gamma_{ac}^r = \delta_c^r K_a + \delta_a^r K_c -C_{acp} S^{rp} + s_{ac}^r.
 \ee
We will need various properties of the closed string propagators. As first pointed out in
\cite{gkmw}, some of these properties are better addressed in the ``big moduli space'' formalism. In this space,
on top of the $s$ complex coordinates $z^a$, one introduces an additional complex coordinate corresponding roughly to the
string coupling constant. In what follows, lower Latin indices run over ${a=1, \dots, s}$, and lower
Greek indices run over the indices of the ``big'' moduli space~${\alpha=0,1, \dots, s}$.

In the big moduli space, a crucial role is played by the projective coordinates for the moduli space,
$X^I$, $I=0, 1, \dots, s$. We recall that a choice of
frame in the topological string is equivalent to a choice of a symplectic basis of
three-cycles in $H_3(X,\IZ)$, $A^I$, $B_I$, $I=0,1, \dots, s$.
Then, the periods of the holomorphic three-form are defined by
\be
\label{omperiods}
X^I=\int_{A^I}\Omega, \qquad \CF_I=\int_{B_I}\Omega.
\ee
The first set of periods defines the projective coordinates, while the second set defines the (projective) prepotential $\CF$ through
\be
\CF_I= {\partial \CF \over \partial X^I}.
\ee
We now introduce
the $(1+s) \times (1+ s)$ matrix (see, e.g.,~\cite{gkmw, hosono, klemm})
\be
\label{chi-def}
\chi^I_\alpha=\begin{cases} D_a X^I & \text{if $\alpha=a=1,2, \dots, s$}, \\
X^I & \text{if $\alpha=0$.}
\end{cases}
\ee
This matrix is invertible, and its inverse will be denoted by $\chi^\alpha_I$. It satisfies
\be
\label{chi-inverse}
\chi^I_\alpha \chi_J^\alpha=\delta^I_J, \qquad \chi^I_\alpha \chi_I^\beta=\delta_\alpha^\beta.
\ee
It is shown in~\cite{hosono} that the quantities
\be
\label{h-def}
h_I= \chi_I^0 + K_a \chi_I^a
\ee
are holomorphic.

Let us now consider the holomorphic limit of the theory, which depends on a choice of frame. This is
specified by a choice of periods $X^I$, $I=0, 1, \dots, s$, and flat coordinates
\be
t^a= {X^a \over X^0}, \qquad a=1, \dots, s.
\ee
In the holomorphic limit, one has~\cite{bcov}
\be
\label{holchrist}
\Gamma_{a b}^{c}\rightarrow {\partial z^{c} \over\partial t^p} {\partial ^2
t^p \over \partial z^{a} \partial z^{b}}, \qquad K_a \rightarrow -\partial_a \log\bigl(X^0\bigr).
\ee
The holomorphic limit of the free energies is obtained by taking the holomorphic limit of the
propagators. In the case of the closed string, this limit has been studied in detail in~\cite{gkkm, gm-multi}.
We will denote the holomorphic limit of the
shifted propagators by calligraphic letters~$\CS^{ij}$,~$\tilde \CS^j$,~$\tilde \CS$.
Based on the results in~\cite{hosono}, it can be shown that these holomorphic
propagators satisfy the equations
\begin{align}
\begin{aligned}
&C_{ijl} \CS^{lk}- s_{ij}^k = -\partial^2_{ij} X^I \chi_I^k,\\
&C_{ijk}\tilde \CS^k - h_{ij} = h_I \partial_{ij}^2 X^I.
\end{aligned}\label{first-const}
\end{align}

Let us now introduce propagators for the real topological string,
following~\cite{walcher-bcov}.\footnote{In~\cite{walcher-tadpole, walcher-bcov},
the real topological string propagators are denoted by $\Delta^b$, $\Delta$.} They are defined by
\be\label{oneleg-prop}
\partial_{\bar a} R^b=\ri \re^K G^{b \bar c} \Delta_{\bar a \bar c}=\Delta_{\bar a}^{~b},
\ee
and
\be
R^a = G^{a \bar b} \partial_{\bar b} R.
\ee
They are both sections of $\CL^{-1}$. One has also the property
 \be
 \Delta_{\bar a \bar b}= -\ri \re^{-K} D_{\bar a} D_{\bar b} R.
 \ee
The propagator $R^k$ is closely related to the Griffiths invariant. To see this, we first note that the HAE (\ref{del-del}) can be written as
\be
\label{hae-griffiths}
\partial_{\bar a} \Delta_{ij}=-C_{ij k} \Delta_{\bar a}^{~k},
\ee
and one has
\be
\label{propa-nice}
C_{ijk} R^k+ \Delta_{ij}= d_{ij},
\ee
where $d_{ij}$ are holomorphic functions. The relation (\ref{propa-nice}) is similar to
(\ref{chris-prop}), and as we will see, it gives a useful expression for the holomorphic limit of the real propagators.
As in the closed string case,
the derivatives of $R^j$, $R$ can be written as polynomials in the propagators. It is convenient
to introduce the tilded real propagators~\cite{al}
\begin{gather}
\tilde R^i= R^i, \qquad
\tilde R= R- K_i R^i.\label{redef-delta}
\end{gather}
Since $R^i$ remains the same, we will omit the tilde. Then, one finds the relations~\cite{al}
\begin{align}
\begin{aligned}
&\partial_i R^j= \delta_i^j \widetilde R + S^{jl} \bigl( C_{ilk} R^k - d_{il} \bigr) - s_{ik}^j R^k + d_i^j, \\
&\partial_i \widetilde R= \bigl( C_{ijk} R^j-d_{ik} \bigr) \widetilde S^k -h_{ij} R^j + d_i,
\end{aligned}\label{add-rels}
\end{align}
where $d_{ij}$, \smash{$d_i^j$} and $d_i$ are holomorphic ambiguities and the $d_{ij}$ are the same quantities appearing in (\ref{propa-nice}).

It is possible to obtain relations between the real topological string propagators and the closed string propagators. Indeed, by using (\ref{griff}),
we find
\begin{align}
\partial_{\bar a} R^b&= \ri \re^K G^{b \bar c} D_{\bar a} D_{\bar c} \overline \CT - \overline C_{\bar a}^{bm} D_m \CT =
\partial_{\bar a} \bigl( \ri \re^K G^{b \bar c} D_{\bar c} \overline \CT \bigr) - \partial_{\bar a} S^{bm} D_m \CT \nonumber\\
&=\partial_{\bar a}\bigl\{ \ri \re^K G^{b \bar c} D_{\bar c} \overline \CT - S^{bm} D_m \CT + S^b \CT \bigr\},\label{RS}
\end{align}
where we have used that
\be
\partial_{\bar a} ( D_m \CT )= G_{\bar a m} \CT,
\ee
since $\CT$ is holomorphic. A similar argument can be applied to $R$, and one finds in this way
\begin{align}
\begin{aligned}
&R^k= - S^{km} \partial _m \CT + \tilde S^k \CT+\ri \re^K G^{k \bar c} D_{\bar c} \overline \CT + r^k, \\
&\tilde R= -\tilde S^m \partial_m \CT + 2 \tilde S \CT + \ri \re^K \bigl( \overline{\CT}- K_b G^{b \bar c} D_{\bar c} \overline \CT \bigr) + r.
\end{aligned}\label{RS-rel}
\end{align}
In these expressions, $r^k$, $r$ are holomorphic functions which
depend explicitly on the domain wall tension $\CT$. For example, one can show that
the functions $r^k$ satisfy
\be
\label{rk-cons}
C_{ijk} r^k = -\partial_{ij}^2 \CT + s^k_{ij} \partial_k \CT - h_{ij} \CT +d_{ij}.
\ee
A similar connection between the real and the closed string propagators is implicit in the expressions for the former
found in~\cite{nw} in the big moduli space.

Let us now consider the holomorphic limit for the real topological string. In order to do this, we have to understand
first the holomorphic limit of Griffiths' invariant. As explained in~\cite{walcher-bcov}, in~a~given frame
one has to choose $\CT$ in such a way that it vanishes at the appropriate base-point, together with all its holomorphic
derivatives. When this is the case, both $\overline \CT$ and $D_{\bar a} \overline \CT$ vanish in the holomorphic limit.
Therefore, the holomorphic limit of Griffiths' invariant, which we will denote by ${\cal D}_{ij}$, is simply given by
\be
\label{hol-grif}
{\cal D}_{ij} = D_i D_j \CT,
\ee
where the covariant derivatives are calculated in the holomorphic limit. In terms of holomorphic propagators, one has
\be\label{holGrif}
{\cal D}_{ij}= \partial_{ij}^2 \CT- \bigl( C_{ijl} \CS^{lk} - s^k_{ij} \bigr) \partial_k \CT + \bigl( C_{ijk} \tilde \CS^{k} - h_{ij} \bigr) \CT.
\ee

Let us now consider the holomorphic limit of the real propagators $R^b$, $\tilde R$ in a given frame. We will use again calligraphic letters and
denote these limits by $\CR^b$, $\tilde \CR$. They satisfy the holomorphic limits of the various equations that we have considered.
For example, from (\ref{RS}) we find
\begin{align}
\begin{aligned}
&\CR^k= - \CS^{km} \partial _m \CT + \tilde \CS^k \CT+ r^k, \\
&\tilde \CR= -\tilde \CS^m \partial_m \CT + 2 \tilde \CS \CT + r.
\end{aligned}\label{hol-limit}
\end{align}
In addition, by taking the holomorphic limit of (\ref{propa-nice}), we obtain
\be
\label{hol-nice}
C_{ijk} \CR^k -d_{ij}=-{\cal D}_{ij}.
\ee

As we have emphasized many times, the original propagators are globally
defined but non-holomorphic. Their holomorphic limit
is frame dependent, and we would like to know how they transform under a change
of frame. This was a crucial ingredient in~\cite{gkkm, gm-multi}
and we will need the corresponding generalization to the real case. We recall that a change of frame is implemented
by a change of symplectic basis in (\ref{omperiods}) and is therefore associated to a symplectic matrix
\be
\Gamma=\begin{pmatrix}A& B \\
 C&D \end{pmatrix},
\label{simp}
\ee
where $A$, $B$, $C$, $D$ are $(1+s) \times (1+s)$ matrices which satisfy
\be
A^{\mathsf T}D-C^{\mathsf T}B= {\bf 1}_{s+1}, \qquad
A^{\mathsf T}C=C^{\mathsf T}A, \qquad
B^{\mathsf T}D=D^{\mathsf T}B,
\label{simple}
\ee
and $ {\bf 1}_{s+1}$ is the identity matrix of rank $s+1$. The matrix $\Gamma$ acts on the periods as
\begin{align}
&X^J_\Gamma= C^{JI}\CF_I + D^J_{~I} X^I, \qquad
\CF_J^\Gamma= A_J^{~I} \CF_I + B_{JI} X^I.\label{periods-G}
\end{align}
One can show~\cite{gkkm, hosono} that the holomorphic shifted propagators in the frame defined by $\Gamma$ are
related to the ones in the original frame by the equations
\begin{align}
\begin{aligned}
&\CS^{kl \Gamma}= \CS^{kl} - \bigl[(C\tau+ D)^{-1} C\bigr]^{IJ} \chi_I^k \chi_J^l, \\
&\tilde \CS^{k\Gamma}=\tilde \CS^k+ \bigl[(C\tau+ D)^{-1} C\bigr]^{IJ} \chi_I^k h_J, \\
& \tilde \CS^\Gamma = \tilde \CS - {1\over 2} \bigl[(C\tau+ D)^{-1} C\bigr]^{IJ} h_I h_J,
\end{aligned}\label{sym-trans}
\end{align}
where \smash{$\chi_I^k$}, $h_I$ were introduced in (\ref{chi-inverse}) and (\ref{h-def}), respectively.

In the case of the real topological string there is an additional ingredient, since
the domain wall tension $\CT$ is in general {\it not} invariant under a change of frame.
The reason is that,
as we change the frame, we want to make sure that the domain wall tension in the new frame, $\CT^\Gamma$,
as well as its holomorphic derivatives,
vanish at the appropriate base-point. The original domain wall tension $\CT$ (or its analytic continuation)
does not satisfy this property. In general,
$\CT^\Gamma$ differs from $\CT$ by a (real) linear combination of periods, i.e.,
\be
\label{Tgamma}
\CT^\Gamma= \CT + \alpha_I X^I + \beta^I \CF_I.
\ee
Note that, since $X^I$, $\CF_I$ satisfy the homogeneous version of the Picard--Fuchs equation, both $\CT$ and $\CT^\Gamma$
solve the defining equation (\ref{inh-pf}) for the domain wall tension.
We also note that the holomorphic functions $r^k$, $r$ appearing in (\ref{RS-rel}) will depend on the choice of frame through their dependence
on $\CT$.

We will need the transformation properties of the real propagators under a change of frame specified by
(\ref{periods-G}) and (\ref{Tgamma}). To derive these, we can first deduce the
transformation properties of the holomorphic Griffiths invariant (\ref{holGrif}), and then recall the relation (\ref{hol-nice}) to derive the transformation
of $\CR^k$. By using the first equation in (\ref{add-rels}) one finally obtains the transformation of $\tilde \CR$. The final result is
\begin{align}
\begin{aligned}
&\CR^{k \Gamma}- \CR^k= \bigl\{ \bigl[(C\tau+ D)^{-1} C\bigr]^{PJ} \partial_P \CT^\Gamma - \beta^J \bigr\} \chi_J^k,\\
&\tilde\CR^{ \Gamma}- \tilde \CR= -\bigl\{ \bigl[(C\tau+ D)^{-1} C\bigr]^{PJ} \partial_P \CT^\Gamma - \beta^J \bigr\} h_J.
\end{aligned}\label{gamma-real}
\end{align}
In deriving these equations it is useful to keep in mind that, since $\CT$ is a section of $\CL$, it is a~homogeneous function of
$X^I$ of degree one, and Euler's theorem gives
\be
\CT=X^I \partial_I \CT.
\ee
As a consequence,
\be
\chi^a_I \partial_a \CT + h_I \CT = \partial_I \CT.
\ee

\subsection{Direct integration}

Let us now rewrite Walcher's HAE in terms of closed and real propagators,
following~\cite{al}. Since the non-holomorphic dependence of the free energies is
contained in the propagators, one obtains the equations
\begin{align}
\begin{aligned}
&{\partial G_\chi \over \partial S^{jk}}=
{1\over 2} \sum\limits_{{\chi_1+\chi_2=\chi-2}\atop{\chi_1,\chi_2\geq 0}} D_j G_{\chi_1} D_k G_{\chi_2} + D_j D_k G_{\chi-2}, \\
& {\partial G_\chi \over \partial R^i}-K_i {\partial G_{\chi} \over \partial \tilde R}=- D_i G_{\chi-1},
\end{aligned}\label{HAE-propas}
 \end{align}
as well as the constraint
\be
{\partial G_\chi \over \partial K_i}=0.
\ee

It is sometimes useful to use HAE for the different ingredients of the total free energy. For example,
the oriented open string amplitudes $F_{g,h}$
satisfy~\cite{walcher-bcov}
\begin{align}
\partial_{\bar a} F_{g,h}={}& {1\over 2} \overline C_{\bar a}^{jk} \sum_{(g_1, h_1) +(g_2, h_2)=(g,h) } D_j F_{g_1, h_1} D_k F_{g_2,h_2}\nonumber\\
&{+}\, {1\over 2} \overline C_{\bar a}^{jk} D_j D_k F_{g-1,h}- R_{\bar a}^j D_j F_{g,h-1},
\end{align}
which leads to
\begin{align}
\begin{aligned}
&{\partial F_{g,h} \over \partial S^{jk}}=
 {1\over 2} \sum_{(g_1, h_1) +(g_2, h_2)=(g,h) } D_j F_{g_1, h_1} D_k F_{g_2,h_2} + {1\over 2} D_j D_k F_{g-1,h},\\
& {\partial F_{g,h} \over \partial R^i}-K_i {\partial F_{g,h} \over \partial \tilde R}=- D_i F_ {g,h-1},
\end{aligned}
 \end{align}
and the constraint
\be
{\partial F_{g,h} \over \partial K_i}=0.
\ee
Similarly, one can write HAE for the non-orientable amplitudes separately.

We can now use the propagators to write explicit expressions for the free energies. Let us first consider the
exceptional case with $\chi=0$. In the case of the genus one free energy $F_1$, we can use the definition of the propagator to integrate (\ref{holan1}) as
\be
D_i F_1 = {1\over2} C_{ijk}S^{jk}- \Bigl( {\chi \over 24}-1 \Bigr) K_i+ f^{(1)}_i(z),
\ee
where \smash{$f^{(1)}_i(z)$} is a holomorphic ambiguity. A similar formula can be obtained for the Klein bottle amplitude,
\be
\label{dk00}
D_i \CK_{0,0}= {1\over2} C_{ijk}S^{jk}- K_i+ k^{(1)}_i(z),
\ee
where \smash{$k^{(1)}_i(z)$} is the corresponding ambiguity.
Finally, for the annulus amplitude one obtains from (\ref{annulus-HAE}), in terms of real propagators,
\be
\label{f02}
\partial_j F_{0,2}= {1\over 2} C_{jkl} R^k R^l -d_{jk} R^k + f_j^{(0,2)}(z),
\ee
where \smash{$ f_j^{(0,2)}(z)$} is the holomorphic ambiguity. By adding all these results, we obtain a
useful formula for the derivative of the total $\chi=0$ amplitude,
\be
\label{djg0}
\partial_j G_0= C_{jlk}S^{kl} -{\chi \over 24} K_j +{1\over 2} C_{jkl} R^l R^k -d_{jk} R^k + g_j ^{(0)}.
\ee
Let us note that, when acting on $G_0$, $D_j =\partial_j$. The result (\ref{djg0})
can be used in the HAE (\ref{HAE-propas}) to calculate all the $G_\chi$ recursively, as polynomials in the
propagators, and up to holomorphic ambiguities. Let us work out the first case, $G_1$. It satisfies the equations
\begin{align}
\begin{aligned}
&{\partial G_1 \over\partial S^{ij}}=\Delta_{ij}= -C_{ijk} R^k + d_{ij}, \\
& {\partial G_1 \over \partial R^j}- K_j {\partial G_1 \over \partial \tilde R}=-D_j G_0.
\end{aligned}
\end{align}
By using (\ref{djg0}), the last equation can be split into two
\begin{align}
\begin{aligned}
&{\partial G_1 \over \partial R^k}= -C_{ijk}S^{ij} -{1\over 2} C_{kij}R^i R^j +d_{kj} R^j + g_k ^{(0)}, \\
&{\partial G_1 \over \partial \tilde R}=-{\chi \over 24}.
\end{aligned}
\end{align}
By integrating these equations, we obtain
\be
G_1= -S^{ij } \bigl( C_{ijk}R^k - d_{ij} \bigr)-{1\over 6} C_{ijk} R^i R^j R^k +{1\over 2} d_{ij} R^i R^j -{\chi \over 24} \tilde R + g_k ^{(0)} R^k + g_1,
\ee
where $g_1$ is a new holomorphic ambiguity.

\subsection{Fixing the holomorphic ambiguity}

Integrating the HAE for the real topological
string requires fixing the holomorphic ambiguity at each
value of $\chi$. As in the closed string case, in order to do this it is helpful to know the behavior
of the real topological string amplitudes at special points in moduli space. Among these, perhaps the
most important one is the conifold locus. It has been known for some time that closed
topological string amplitudes
have a universal behavior at this locus, given by the $c=1$ string~\cite{gv-conifold}. Let us
assume for simplicity that there is a single
modulus in the geometry, and let $t_c$ be a vanishing flat coordinate at the conifold locus.
Then, in the conifold frame, the closed
string amplitudes have the following behavior:
\be
\label{gap}
\CF_g^c (t_c)= {B_{2g} \over 2g (2g-2)}t_c^{2-2g} + \CO(1),
\ee
where $B_{2g}$ are Bernoulli numbers, and we have normalized $t_c$ appropriately
\big(here and in the following we use inhomogeneous free energies,
which differ from the homogeneous ones in an~appropriate power of $X^0$\big).
According to (\ref{gap}), there are no
subleading poles in the free energy near the conifold point. This gap condition was
emphasized in~\cite{hk06}
and it was later realized that, in many toric CY manifolds, it makes it possible to fix the
holomorphic ambiguity at all genera~\cite{hkr}.

The behavior of the real topological string amplitudes near the conifold point was studied in~\mbox{\cite{kw-localp2,kw,walcher-tadpole, walcher-bcov}}.
It was found that all real topological string amplitudes $\CF_{g,h}$, $\CK_{g,h}$ with $h\not=0$ were regular at the
conifold point, except for $\CK_{g,0}$, which in the conifold frame and near the conifold point behaves as
\be
\label{k-gap}
\CK^c_{g-1,0}(t_c)= {\Psi_g \over t_c^{2g-2}}+ \CO(1).
\ee
A proposal for the value of the coefficient $\Psi_g$ was made in~\cite{kw}, and
it is given by
\be\label{gap-real}
\Psi_g={1 \over 2^{2g+1} g(g-1)}\bigl\{ -\bigl(2^{2g}-1\bigr) B_{2g}+ g E_{2g-2}\bigr\},
\ee
where $B_{2g}$ are Bernoulli numbers and $E_{2g-2}$ are Euler numbers.
We conclude that $\CG_\chi$ has a gap behavior at the conifold
for $\chi$ even,
and is regular for $\chi$ odd. This behavior helps in fixing the holomorphic ambiguity, but in contrast to what happens in the closed
string case, it does not fix it completely, even in the local case, and one has to use further information.
We will discuss this in more detail in Section~\ref{sec-p2}.

 \section{Trans-series solutions and resurgent structure }
\label{sec:trans-series-res-structure}

 Our discussion of the real topological string has so far focused on its perturbative expansion.
 Following~\cite{cesv2,cesv1,cms, gkkm,gm-multi}, we would like
 to consider now {\it trans-series solutions} to Walcher's HAE. These will be used to obtain information on the
 non-perturbative sector of the real topological string, and in particular to find explicit multi-instanton amplitudes.

 \subsection{Master equation and warm-up example}

 In order to obtain a trans-series solution, we have to re-express first the HAE as a ``master equation'' for the full free energy $G$. Due to
 the special r\^ole of the $G_\chi$ with $-2\le \chi\le 0$, we have to consider four different generating series, which we define as
 \begin{alignat}{3}
 \begin{aligned}
& G^{(0)}= \sum_{\chi \ge -2} G_\chi g_s^\chi, \qquad&&
 \widetilde G^{(0)} = \sum_{\chi \ge -1} G_\chi g_s^{\chi},& \\
& \widehat G^{(0)}= \sum_{\chi \ge 0} G_\chi g_s^{\chi}, \qquad&&
 \overline G^{(0)} = \sum_{\chi \ge 1} G_\chi g_s^{\chi}.&
 \end{aligned}\label{four-series}
 \end{alignat}
 The superscript \smash{$^{(0)}$} indicates that these are all perturbative amplitudes, and the overline in the last series should not be confused
 with complex conjugation. We will first consider the following trans-series ansatz to solve the HAE,
 \be
 \label{one-par-ts}
 G=\sum_{\ell \ge 0} \CC^\ell G^{(\ell)},
 \ee
 where $\CC$ is a formal parameter keeping track of the exponential weight, and
 \be
 \label{inst-gser}
 G^{(\ell)}= \sum_{n \ge 0} \re^{- \ell \CA/g_s} G_n^{(\ell)} g_s^n, \qquad \ell \ge 1.
 \ee
We will often refer to $\CA$ as an ``instanton action'' and to the \smash{$G^{(\ell)}$} as $\ell$-th instanton amplitudes (a more
general ansatz will be considered below, in Section~\ref{sec-multinsts}).
Our goal is to find explicit expressions for the amplitudes \smash{$G^{(\ell)}$}.
We define the trans-series analogue of \smash{$\widetilde G^{(0)}$} as
\be
\widetilde G= \widetilde G^{(0)}+ \sum_{\ell \ge 1} \CC^\ell G^{(\ell)},
\ee
and similarly for \smash{$\widehat G$} and $\overline G$.

 Before working out the general trans-series solution, it is instructive to solve a simple, particular case by hand, as it
 was done originally in~\cite{cesv2, cesv1}. The simplest
 situation is the one considered in~\cite{gm-multi} for closed topological strings, namely
 toric or local CY manifolds with a single modulus.
 In the local case, as first noted in~\cite{kz}, the HAE simplify substantially:
the holomorphic limit of~$K_a$ vanishes, and the closed string propagators \smash{$\tilde \CS^k$}, \smash{$\tilde S$} can be
set to zero by an appropriate choice of the holomorphic functions appearing in the equations
(\ref{ders-tilded}). Similarly, one
can set to zero the holomorphic real topological string propagator $\tilde \CR$ by choosing $h_{ij}=d_i=0$, as required by the
second equation in (\ref{add-rels}). Therefore, if we are interested in the end in calculating holomorphic quantities, we can
set \smash{$\tilde S^k$}, \smash{$\tilde S$} and \smash{$\tilde R$} to zero from the very beginning. Since we want to consider a
 single modulus case, there are only two non-zero propagators, namely $S^{zz}$ and $R^z$. Taking all this into account, the HAE (\ref{HAE-propas}) simplify to
 \begin{align}
 \begin{aligned}
& { \partial G_\chi \over \partial S^{zz}} = \frac{1}{2}\sum\limits_{{\chi_1+\chi_2=\chi-2}\atop{\chi_1,\chi_2\geq 0}} D_z G_{\chi_1} D_z G_{\chi_2} + D^2_z G_{\chi-2},\\
& { \partial G_{\chi} \over \partial R^z}= - D_z G_{\chi-1}.
\end{aligned}\label{hae-sd}
\end{align}
We first write down master equations for the perturbative series, and we then postulate that these equations are valid for
their trans-series counterparts. It is easy to see that one obtains in this way
 \begin{align}
 \begin{aligned}
& { \partial \overline G \over \partial S^{zz}}= {g_s^2 \over 2} \bigl( D_z \widehat G \bigr)^2+ g_s^2 D_z^2 \widetilde G,\\
 & { \partial \overline G \over \partial R^z }= - g_s D_z \widehat G.
\end{aligned}
 \end{align}
 We will focus in this section on the one-instanton amplitude. It satisfies the linearized master equations
 \begin{align}
\begin{aligned}
& { \partial G^{(1)} \over \partial S^{zz}}= g_s^2 D_z \widehat G^{(0)} D_z G^{(1)}+ g_s^2 D_z^2 G^{(1)} ,\\
& { \partial G^{(1)} \over \partial R^z }= - g_s D_z G^{(1)}.
\end{aligned}
 \end{align}
 We now plug in these equations the ansatz (\ref{inst-gser}) and we solve order by order in $g_s$.
 We find that, as in the closed topological string case~\cite{cesv2, cesv1}, $\CA$ is independent of
 the propagators. We will assume that the relevant instanton actions are, up to an overall constant factor,
 periods of the holomorphic 3-form (in the B-model picture). In the standard topological string, they actually
 belong to the {\it integral} lattice of periods,
 as emphasized in~\cite{gkkm}. However, as we will see in Section~\ref{sec-bc}, half-integral periods
 also play a role in the real topological string.

 The next order gives the following equations for \smash{$G_0^{(1)}$}:
 \begin{align}
 \begin{aligned}
 &{\partial G_0^{(1)} \over \partial S^{zz}}= (\partial_z \CA)^2 G_0^{(1)},\\
 & {\partial G_0^{(1)} \over \partial R^z }= \partial_z \CA \, G_0^{(1)}.
\end{aligned}
 \end{align}
 This can be immediately integrated to obtain
 \be
 G_0^{(1)}= \exp \bigl( (\partial_z \CA)^2 S^{zz} + \partial_z \CA \, R^z\bigr) g_0^{(1)},
 \ee
 where \smash{$g_0^{(1)}$} is independent of the propagators, and it is the trans-series counterpart of the holomorphic ambiguity.
 As first explained in~\cite{cesv2,cesv1}, this ambiguity is fixed by evaluating the trans-series in a special
 frame, called the $\CA$-frame. This is a frame in which $\CA$ is (up to possible overall constant factors)
 one of the $A$-periods. In this frame we impose
 a fixed, simple form for the multi-instanton amplitude, which we will call, as in~\cite{gkkm,gm-multi}, a {\it boundary condition}.
 We will discuss different boundary conditions in Section~\ref{sec-bc}. In the case at hand, as we will explain below, the boundary condition
 is simply
 \be
 \label{bc-ex}
 \CG_{0,\CA}^{(1)}=1,
 \ee
 where the subscript $\CA$ means evaluation in the $\CA$-frame. On the other hand, from the above expression we find
 \be
 \CG_{0,\CA}^{(1)} = \exp \bigl( (\partial_z \CA)^2 \CS^{zz}_\CA + \partial_z \CA \, \CR^z_\CA \bigr) g_0^{(1)},
 \ee
 and this fixes \smash{$g_0^{(1)}$}. We obtain in the end
 \be
 \label{eq:oneInstFromHAE}
 G_0^{(1)}= \exp \bigl( (\partial_z \CA)^2 (S^{zz}-\CS^{zz}_\CA) + \partial_z \CA \, (R^z-\CR^z_\CA) \bigr).
 \ee
 This procedure can be pushed to calculate $G^{(1)}$ at higher orders. At next to leading order we find, for example,
\begin{align}
G^{(1)}_1={}& \exp\bigl[(\partial_z\mathcal{A})^2(S^{zz}-\mathcal{S}_{\mathcal{A}}^{zz}) + \partial_z\mathcal{A}(R^z-\mathcal{R}_{\mathcal{A}}^z)\bigr]\frac{ \partial_z\mathcal{A} (S^{zz}-\mathcal{S}_{\mathcal{A}}^{zz})}{6}\nonumber\\
&{\times}\,\bigl[4C_{zzz}(\partial_z\mathcal{A})^2(S^{zz}-\mathcal{S}_{\mathcal{A}}^{zz})^2 +
+6C_{zzz}(S^{zz}-\mathcal{S}_{\mathcal{A}}^{zz})(1+(\partial_z\mathcal{A})R^{z})\nonumber\\
&\hphantom{{\times}\bigl[}+3C_{zzz}\bigl((R^z-\mathcal{R}_{\mathcal{A}}^z)^2+2\mathcal{S}_{\mathcal{A}}^{zz} +2(R^z-\mathcal{R}_{\mathcal{A}}^z)\mathcal{R}_{\mathcal{A}}^z+(\mathcal{R}_{\mathcal{A}}^z)^2\bigr)+6k_z^{(1)}(z)\bigr].\label{eq:oneInstnexttoleading}
\end{align}
The above expression is already quite complicated and pushing the calculation to higher orders immediately becomes cumbersome, although there is no conceptual difficulty. In addition,
 the meaning of the multi-instanton amplitudes is not clear in this language.
 For this reason, we will adopt the operator formalism
 first discussed in~\cite{codesido-thesis, coms}.

\subsection{Operator formalism}

In order to solve the HAE for the trans-series of the real topological string,
we generalize the operator formalism introduced in~\cite{gkkm,gm-multi}. We have to
introduce first a generalized $U(1)$ covariant derivative
\be
\mathfrak{d}_\alpha = (\mathfrak{d}_0, \mathfrak{d}_a),
\ee
where
\be
\mathfrak{d}_0= -g_s {\partial \over \partial g_s}, \qquad \mathfrak{d}_a = \mD_a- K_a g_s {\partial \over \partial g_s}.
\ee
Here, $\mD_a$ acts on a function of $z^a$, the closed and real propagators $S^{ab}$, $\tilde S^a$, $\tilde S$, $R^i$, $\tilde R$, and~$K_a$,~as%
\begin{gather}
\mD_a f= {\partial f\over \partial z^a} + \partial_a S^{de} {\partial f \over \partial S^{de}} +\partial_a \tilde S^{c} {\partial f \over \partial \tilde S^{c}} +
\partial_a \tilde S {\partial f \over \partial \tilde S} + \partial_a R^{c} {\partial f \over \partial R^{c}} +
\partial_a \tilde R {\partial f \over \partial \tilde R}+ \partial_a K_{c} {\partial f \over \partial K_c},\!\!\!
\end{gather}
where the derivatives of the propagators
are computed with the rules (\ref{ders-tilded}) and (\ref{add-rels}).
The basic operator of the formalism is defined in such a way that, in the holomorphic limit,
it will become a derivative with respect to the
flat projective coordinates $X^I$. One defines~\cite{gkkm}
\begin{align}
\begin{aligned}
&T^j=g_s \bigl\{ - \CA \bigl(\tilde S^j - \tilde \CS^j_\CA \bigr) + \partial_m\CA \bigl(S^{mj}-\CS_\CA ^{mj}\bigr) \bigr\},\\
&T^0= g_s \bigl\{ 2 \CA \bigl( \tilde S- \tilde \CS_\CA\bigr) - \partial_m\CA \bigl(\tilde S^m- \tilde \CS^m_\CA\bigr) \bigr\}- K_j T^j.
\end{aligned}
\label{exTs}
\end{align}
These are used to construct the operator $\oD$ as
\be
\label{eq:op-D}
\oD = T^\alpha \mathfrak{d}_\alpha=T^j \mathfrak{d}_j + T^0 \mathfrak{d}_0,
\ee
and we will sometimes decompose it as
\be
\oD= \oD_0 + \oD_1,
\ee
where
\be
\oD_0= T^0 \mathfrak{d}_0, \qquad \oD_1= T^j \mathfrak{d}_j.
\ee
It is also convenient to introduce
\be
\tilde T^0= T^0 + K_j T^j =g_s \bigl\{ 2 \CA ( S- \CS_\CA) - \partial_m\CA \bigl(\tilde S^m- \tilde \CS^m_\CA\bigr) \bigr\}.
\ee
We have the crucial relations~\cite{gkkm}
 \begin{align}
 \begin{aligned}
 &\mathfrak{d}_i T^j= - \Gamma_{ik}^j T^k - T^0 \delta_i^j, \\
 &\mathfrak{d}_i T^0=0,
 \end{aligned}\label{gen-ts}
 \end{align}
and we note that
\be
\mathfrak{d}_i \tilde T^0=-K_i \tilde T^0 - \bigl(C_{ijk} \tilde S^k - h_{ij} \bigr) T^j.
\ee
We also introduce derivative operators with respect to the propagators, just as in~\cite{gkkm}:
\begin{align}
\begin{aligned}
&\delta^S_{ij}= {1\over g_s^2} \biggl( {\partial \over \partial S^{ij}}- K_{(i } {\partial \over \partial \tilde S^{j)}} +{1 \over 2} K_i K_j {\partial \over \partial \tilde S } \biggr), \\
&\delta^S_{0i}= -{1\over g_s^2} \biggl( {\partial \over \partial \tilde S^i}-K_i {\partial \over \partial \tilde S}\biggr), \\
&\delta^S_{00}= {1\over 2 g_s^2} {\partial \over \partial \tilde S}.
\end{aligned}
\end{align}
and we define
\be
\omega_S = T^i T^j \delta^S_{ij} +T^0 T^i \delta^S_{0i}+ T_0^2 \delta^S_{00},
\ee
as well as the operator
\be
\label{eq:op-W}
\mW= \omega_S -\oD \widehat G \, \oD,
\ee
where $\widehat G= \widehat G^{(0)}$ is one of the formal series in (\ref{four-series}).

We now write the HAE in terms of these operators. The equation for the dependence on the closed string propagators is
similar to what is obtained in the closed string case,
\be
\mW \overline G = \oD G_0 \oD \widehat G -{1 \over 2} \bigl(\oD \widehat G \bigr)^2 + \oD^2 \widetilde G.
\ee
In this equation, $\oD^2 \widetilde G$ is a series in $g_s$ and its first term has to be understood as
\be
\label{d2g1}
 \oD^2 \biggl( {G_{-1} \over g_s} \biggr)={1\over g_s} T^i T^j \mD_i \mD_j G_{-1}= {1\over g_s} T^i T^j \Delta_{ij}.
 %= -{1\over g_s} T^i T^j (C_{ijk} R^k-d_{ij}).
\ee

Let us now consider the second equation in (\ref{HAE-propas}), involving the open string propagators.
By considering different powers of $K_i$, we obtain two different equations
\be
\ba
{\partial G_\chi \over \partial R^i}= -\mD_i G_{\chi-1},\qquad
{\partial G_\chi \over \partial \tilde R}= (1-\chi) G_{\chi-1},
\ea
\ee
which in terms of generating functions read
\be
\label{R-eqns}
{\partial \overline G \over \partial R^i}= -g_s \mD_i \widehat G,\qquad {\partial \overline G \over \partial \tilde R}= g_s \mathfrak{d}_0 \widehat G.
\ee
This suggests introducing the operator
\be
\label{mRdef}
\mR={1\over g_s} \biggl( T^j {\partial \over \partial R^j} - \tilde T^0 {\partial \over \partial \tilde R} \biggr).
\ee
This is the new ingredient in the operator formalism that we need in order to study the real
topological string. As in~\cite{gkkm}, we will require the operators to have zero charge with respect to the $U(1)$ connection on $\CL$. This is why we have included a factor of $g_s$ in the right-hand side of~(\ref{mRdef}) \big(since $R^i$, $\tilde R$ have charge one\big).
Then, the equations (\ref{R-eqns}) can be written as
\be
\label{RG}
\mR \overline G= - \oD \widehat G.
\ee

We then have three operators: $\mR$, $\mW$ and $\oD$, and we want to understand their commutator algebra.
One can immediately calculate
\begin{align}
\begin{aligned}
&\biggl[ {1\over g_s} {\partial \over \partial R^k}, \mathfrak{d}_i \biggr]= {1\over g_s} \bigl( S^{jl}C_{il k} - s^j_{ik}-K_i \delta_k^j \bigr) {\partial \over \partial R^j} + {1\over g_s} \bigl( \tilde
S^{j}C_{ij k} -h_{ik} \bigr) {\partial \over \partial \tilde R}, \\
&\biggl[ {1\over g_s} {\partial \over \partial \tilde R}, \mathfrak{d}_i\biggr]={1\over g_s} \biggl( {\partial \over\partial R^i} -K_i {\partial \over \partial \tilde R} \biggr),
\end{aligned}
\end{align}
and from this result one easily verifies that $\mR$ and $\oD$ commute:
\be\label{rd-compact}
[ \mR, \oD ]=0.
\ee

To compute the other commutators, some extra work is needed. To obtain the commutator between $\mW$ and $\oD$, we have to
calculate
\begin{align}
\biggl[ {1\over g_s^2} {\partial \over \partial S^{ij}}, \mathfrak{d}_k \biggr]={}&
{-}{1\over g_s^2}\biggl( \Gamma_{i k}^m {\partial \over \partial S^{mj}}+ \Gamma_{j k}^m {\partial \over \partial S^{mi}} \biggr)\nonumber\\
&{+}\, {1\over 2} \bigl(C_{ikb} \tilde S^b - h_{ik} \bigr) {1\over g_s^2} {\partial \over \partial \tilde S^{j}}+ {1\over 2} \bigl(C_{jkb} \tilde S^b - h_{jk} \bigr) {1\over g_s^2} {\partial \over \partial \tilde S^i}\nonumber\\
&{-}\, {1\over 2 g_s^2 } C_{ik m} K_j {\partial \over \partial K_m}-{1\over 2 g_s^2} C_{jk m} K_i {\partial \over \partial K_m} \nonumber\\
&{+}\, {1\over 2 g_s^2} ( C_{kjm} R^m -d_{kj} ) {\partial \over \partial R^i}+{1\over 2 g_s^2} ( C_{kim} R^m -d_{ki} ) {\partial \over \partial R^j},
\end{align}
as well as
\begin{gather}
\begin{aligned}
&\biggl[ {1\over g_s^2} {\partial \over \partial \tilde S^i}, \mathfrak{d}_j \biggr]= {2 \over g_s^2} {\partial \over \partial S^{ij}}
- {1\over g_s^2}\Gamma_{ij}^m
{\partial \over \partial \tilde S^m} + \bigl(C_{ijk} \tilde S^k- h_{ij} \bigr) {1\over g_s^2}{\partial \over \partial \tilde S}- {1\over g_s^2} C_{ijm} {\partial \over \partial K_m}\\
& \hphantom{\biggl[ {1\over g_s^2} {\partial \over \partial \tilde S^i}, \mathfrak{d}_j \biggr]=}{}
 +{1\over g_s^2} ( C_{ijp} R^p- d_{ij} ) {\partial \over \partial \tilde R},\\
& \biggl[ {1\over g_s^2} {\partial \over \partial \tilde S}, \mathfrak{d}_i \biggr] = {2 \over g_s^2} \biggl( {\partial \over \partial \tilde S^i}- K_i {\partial \over \partial \tilde S} \biggr).
\end{aligned}
\end{gather}
When the real propagators are set to zero, one recovers the results of~\cite{gkkm}.
Let us introduce
\be
\delta_j^R= {\partial \over \partial R^j} - K_j {\partial \over \partial \tilde R}.
\ee
From here, one obtains
\begin{align}
\begin{aligned}
&\bigl[\delta^S_{ij}, \mathfrak{d}_k \bigr]= - \Gamma^m_{ik} \delta^S_{jm}- \Gamma^m_{jk} \delta^S_{im}
+{1\over 2g_s^2} ( C_{kim} R^m -d_{ki} )\delta_j^R +{1\over 2g_s^2} ( C_{kjm} R^m -d_{kj} ) \delta^R_i, \\
&\big[\delta^S_{0i}, \mathfrak{d}_j \big]= -2 \delta^S_{ij} -\Gamma^m_{ij} \delta^S_{0m}+ {1\over g_s^2}
C_{ijm} {\partial \over \partial K_m}, \\
&\big[\delta^S_{00}, \mathfrak{d}_i \big]= - \delta^S_{0i},
\end{aligned}
\end{align}
which leads to
\be
\label{simple-comm}
 [ \omega_S, \mathfrak{d}_k ]={1\over g_s}T^i ( C_{ipk} R^p - d_{ik} ) \mR +
{1\over g_s^2} T^0 T^i C_{ikm} {\partial \over \partial K_m}.
\ee
In addition, we have from~\cite{gkkm}
\be\label{omT}
\omega_S (T^\alpha)= T^\alpha \oD\biggl({\CA \over g_s} \biggr),
\ee
and one can also deduce
\be\label{magic-comm}
[\omega_S, \oD] = \oD \biggl( {\CA \over g_s} \biggr) \oD +{1\over g_s}T^i T^k ( C_{ipk} R^p - d_{ik} ) \mR+ {1\over g_s^2} T^0 T^i T^j C_{ijm} {\partial \over \partial K_m}.
\ee
We conclude that
\be\label{magic-comm2}
[\mW, \oD] = \oD H_0 \oD + {1\over g_s}T^i T^k ( C_{ipk} R^p - d_{ik} ) \mR+
 {1\over g_s^2} T^0 T^i T^j C_{ijm} {\partial \over \partial K_m},
\ee
where
\be\label{H0def}
H_0= {\CA \over g_s} + \oD \widehat G.
\ee
This is the analogue of the functional $\CG$ introduced in~\cite{coms,gkkm, gm-multi}.

It is convenient to write (\ref{magic-comm2}) in a slightly different form. We introduce the quantity
\be
\label{H1def}
H_1= -\mD_i \CA \bigl(R^i- \CR^i_\CA \bigr) +\CA \bigl( \tilde R- \tilde \CR_\CA \bigr).
\ee
By using the result of~\cite{gkkm}
\be
\partial_{ij}^2 \CA = -\partial_l \CA \bigl(C_{ijm} \CS^{ml}_\CA - s^l_{ij} \bigr)+ \CA \bigl( C_{ijm} \tilde \CS^{m}_\CA - h_{ij} \bigr),
\ee
one can show that
\be
\label{dh-compact}
\oD H_1= -{1\over g_s} T^i T^k ( C_{ikp} R^p- d_{ik} ),
\ee
therefore we can write (\ref{magic-comm2}) as
\be
\label{magic-comm21}
[\mW, \oD] = \oD H_0 \, \oD -\oD H_1 \, \mR+
 {1\over g_s^2} T^0 T^i T^j C_{ijm} {\partial \over \partial K_m}.
\ee
There are two other properties of $H_1$ which will be needed:
\be
\label{d2gh}
\oD^2 \biggl( { G_{-1} \over g_s} \biggr)= \oD H_1, \qquad \mR \oD G_0= -\oD H_1.
\ee

Finally, we want to calculate the commutator of $\mW$ and $\mR$. By using that
\be
[\omega_S, \mR]= \oD \biggl( {\CA \over g_s} \biggr) \mR,
\ee
one finds
\be
 [ \mW, \mR ]= \oD \biggl( {\CA \over g_s} \biggr) \mR- \bigl( \oD H_1 + \oD^2 \widehat G \bigr) \oD.
\ee

One of the key aspects of the operator formalism constructed in~\cite{coms,gkkm,gm-multi} is that, in the holomorphic limit,
$\oD$ becomes indeed a derivative operator with respect to the flat coordinates of the~CY. The more general
case was analyzed in~\cite{gkkm}. Let us write the
action $\CA$ as an integer linear combination of periods,
\be
\label{action-periods}
\kappa^{-1}\CA= c^J \CF_J + d_J X^J.
\ee
In this formula, $\kappa$ is a universal normalization constant which depends on a choice of normalization for the string coupling constant, see~\cite{gkkm} for a detailed discussion. We will set it to one in most of the formulae that follow.
Let $f$ be a homogeneous function of degree $n$ in the big moduli space coordinates. Then, in the holomorphic limit, one has~\cite{gkkm}
\be
\label{hol-D}
\oD (g_s^{-n} f)\rightarrow g_s^{-n+1} c^{I} {\partial f \over \partial X^I}.
\ee

\subsection{The one-instanton amplitude}\label{subsec:one-inst-amp-op}

We are now ready to use the formalism above to obtain the one-instanton amplitude in closed form. It satisfies the
linearized HAE, which in the operator language that we have just developed is given by
 \begin{align}
 \begin{aligned}
 &\mW G^{(1)}= \oD^2 G^{(1)},\\
 & \mR G^{(1)} =- \oD G^{(1)}.
 \end{aligned}\label{oneinst-op}
 \end{align}
 To solve this equation, we introduce an exponential ansatz, as in~\cite{gm-multi},
 \be
 G^{(1)}= \re^{-\Phi}.
 \ee
 In terms of $\Phi$, the equations (\ref{oneinst-op}) read
 \begin{align}
 \begin{aligned}
 &\mW \Phi = \oD^2 \Phi- ( \oD \Phi )^2, \\
 & \mR \Phi =- \oD \Phi.\end{aligned}\label{phi-eqns}
 \end{align}
 To solve these equations, we consider
 \be
 \label{eq:def-H}
 H= H_0 + H_1,
 \ee
 where $H_{0,1}$ were defined in (\ref{H0def}) and (\ref{H1def}), respectively.
One can easily show that it satisfies the equations
 \begin{align}
 &\mW H= \oD^2 H, \qquad
 \mR H= - \oD H. \label{J-eqns}
 \end{align}
 In proving the first equation, it is useful to note that
\be
\omega_S ( \oD G_0 )= \oD \biggl( {\CA \over g_s}\biggr) \oD G_0 + \oD^2 \biggl({\CA \over g_s} \biggr).
\ee

We now claim that (\ref{phi-eqns}) is solved by
\be\label{phi-exp}
\Phi= \mO_\lambda H,
\ee
where $\mO_\lambda $ is the operator
\be\label{O-def}
\mO_\lambda=\sum_{k \ge 1} {(-\lambda)^{k-1} \over k!} \oD^{k-1} = {1\over \lambda \oD}\bigl(1- \re^{- \lambda \oD}\bigr)= {1\over \lambda} \int_0^{\lambda } \re^{-u \oD} \rd u,
\ee
and we will have to set
\be
\lambda=2.
\ee
The check that (\ref{phi-exp}) satisfies the second equation in (\ref{phi-eqns}) follows immediately from the second equation
in (\ref{J-eqns}) and the commutation relation (\ref{rd-compact}). To prove the first equation, we have to work harder. We first
obtain the commutator of $\mW$ with $\mO_\lambda$. As in
\cite{gkkm,gm-multi}, we calculate the commutator of $\mW$ with $\re^{-u \oD}$ with the help of Hadamard's lemma,
\be
\re^A B \re^{-A}= \sum_{n \ge 0} {1\over n!} [A,B]_n,
\ee
where the iterated commutator $[A,B]_n$ is defined by
\be
[A, B]_n=[A, [A, B]_{n-1}], \qquad [A, B]_0=B.
\ee
In our case, we have the simple result that
\be
[\oD, \mW]_n =- ( \oD^n H_0 ) \oD + ( \oD^n H_1 ) \mR.
\ee
We conclude that
\be
\mW\, \re^{-u \oD}= \re^{-u\oD } \Biggl( \mW - \sum_{k \ge 1} {u^k \over k!} \bigl( \bigl( \oD^k H_0 \bigr) \oD -\bigl( \oD^k H_1 \bigr) \mR \bigr) \Biggr).
\ee
By using now the integral formula for $\mO_\lambda $ in (\ref{O-def}), we find
\be
\mW \mO_\lambda = \mO_\lambda \mW -{1\over \lambda} \int_0^{\lambda } \rd u \, \re^{-u \oD} \bigl\{ \bigl[ \bigl(\re^{u \oD}-1\bigr) H_0 \bigr] \oD-\bigl[ \bigl(\re^{u \oD}-1\bigr) H_1 \bigr] \mR \bigr\}.
\ee
From this, and by taking into account the second equation in (\ref{J-eqns}), one obtains
\be
\mW \Phi= \oD^2 \Phi -{1\over \lambda } \int_0^{\lambda } \bigl[ \bigl(1 -\re^{-u \oD}\bigr) H \bigr] \bigl[\re^{-u \oD} \oD H \bigr].
\ee
On the other hand, we have that
\be
\oD \Phi= \sum_{k \ge 1} {(-\lambda)^{k-1} \over k!} \oD^{k} H =
{1\over \lambda } \int_0^{\lambda } \re^{- u \oD} \oD H \rd u.
\ee
Its square can be computed as
\begin{align}
(\oD \Phi )^2&= {1\over \lambda^2 } \int_0^{\lambda } \re^{- u \oD} \oD H \, \rd u \int_0^{\lambda} \re^{- v \oD} \oD H\, \rd v=
 {2 \over \lambda^2 } \int_0^{\lambda} \re^{- u \oD} \oD H \, \rd u \int_0^{u } \re^{- v \oD} \oD H \rd v \nonumber\\
 &={2 \over \lambda^2 } \int_0^{\lambda} \bigl[\re^{- u \oD} \oD H \bigr]\bigl[ \bigl(1- \re^{-u\oD}\bigr) H \bigr] \,\rd u.
 \end{align}
In going to the last line, instead of integrating the symmetric function in $v$, $u$ over the square ${[0, \lambda]^2}$, we integrated it over the triangle below the diagonal, and multiplied the result by two. We conclude that
\be
{1\over \lambda } \int_0^{\lambda } \re^{-u \oD} \bigl\{ \bigl[ (\re^{u\oD}-1) H \bigr] \oD H \bigr\} \, \rd u
={ \lambda \over 2} (\oD \Phi )^2.
\ee
Therefore,
\be
\mW \Phi= \oD^2 \Phi-{ \lambda \over 2} (\oD \Phi )^2,
\ee
and by setting $\lambda=2$ we obtain the sought-for result.

Since the solution to the HAE is expressed in terms of $H$, we have to calculate its holomorphic limit $\CH$. To do
this it is convenient to introduce a modified prepotential or $\chi=-2$ free energy as in~\cite{gkkm,gm-multi}, defined by
\be
c^I { \partial \widetilde \CG_{-2} \over \partial X^I}=\CA= c^I \CF_I+ d_I X^I.
\ee
It differs from the usual one by quadratic terms in $X^I$. Then, the holomorphic limit of $H_0$ is
\be
\CH_0 = g_s^{-1} c^I \partial_I \widetilde \CG_{-2} + \sum_{\chi \ge 0} c^I \partial_I \CG_\chi g_s^{\chi+1}.
\ee
We should also note that, as found in~\cite{gkkm}, in the compact CY case, the genus one closed string free energy $\CF_1$ appearing here is in fact given by
\be
\CF_1-\Bigl( {\chi \over 24}-1 \Bigr)\log X^0.
\ee
The only thing that remains to do is to calculate the holomorphic limit of $H_1$, which will be denoted by $\CH_1$. By using
(\ref{gamma-real}), we obtain
\be
\CH_1= \bigl[(C\tau+ D)^{-1} C\bigr]^{PJ} \partial_P \CT^\Gamma \bigl( \partial_i \CA \chi^i_J +\CA h_J \bigr) - \beta^J\bigl( \partial_i \CA \chi^i_J +\CA h_J \bigr),
\ee
where
\be
c^I= C^{1I}, \qquad d_I = D^1_{~I},
\ee
since we regard \smash{$\CA= X^1_\CA$} as the first coordinate in the $\CA$-frame. We now use
\be
 \partial_i \CA \chi^i_J +\CA h_J= c^P \tau_{PJ}+ d_J
 \ee
 to obtain
 \be
 \CH_1= c^P \partial_P \CT^\Gamma -\beta^J c^P \tau_{PJ} - \beta^J d_J= c^P \partial_P \bigl( \CT^\Gamma -\beta^J \CF_J \bigr) - \beta^J d_J,
 \ee
 or equivalently,
 \be
 \CH_1= c^P \partial_P \bigl( \CT+ \alpha_J X^J \bigr) - \beta^J d_J.
 \ee
This can be used to define a new $\tilde \CG_{-1}$ as
\be
\CH_1= c^P \partial_P \tilde \CG_{-1},
\ee
which differs from $\CG_{-1}= \CT$ in linear terms in the $X^I$. We then consider a redefined total free energy,
\be
\tilde \CG= {1\over g_s^2} \tilde \CG_{-2} + {1\over g_s} \tilde \CG_{-1}+ \sum_{\chi \ge 0} \CG_\chi g_s^{\chi},
\ee
in such a way that
\be
\CH= g_s c^I \partial_I \tilde \CG.
\ee
We conclude that the holomorphic limit of $\Phi$ is
\be
{1\over 2} \bigl(\tilde \CG\bigl(X^I\bigr)- \tilde \CG\bigl(X^I-2 g_s c^I\bigr) \bigr),
\ee
and the holomorphic one-instanton amplitude is given by
\be
\label{eq:One-Inst-HAE-Sol}
\CG^{(1)}= \exp\biggl[{1\over 2} \bigl(\tilde \CG\bigl(X^I-2 g_s c^I\bigr)-\tilde \CG\bigl(X^I\bigr) \bigr) \biggr].
\ee
This is very similar to the result in~\cite{gkkm,gm-multi}: the one-instanton amplitude is obtained by
shifting the flat coordinates by an integer multiple of the string coupling constant (up to normalization). To illustrate the above result let us list the first few orders in the $g_{s}$ expansion of the one-instanton amplitude \eqref{eq:One-Inst-HAE-Sol}. We have
\begin{align}
\mathcal{G}^{(1)} = \text{e}^{-\mathcal{A}/g_s}\exp\bigl[c^{I} c^{J} \tau_{IJ}-\mathcal{H}_1\bigr]\biggl(1+g_{s}\Upsilon_1 + g_{s}^2\biggl(\Upsilon_2+\frac{1}{2}\Upsilon_1^2\biggr) + \cdots\biggr),
\end{align}
where
\begin{align}
 &\Upsilon_1= -\frac{2}{3}c^{I} c^{J} c^{K} C_{IJK} + c^{I} c^{J}\frac{\partial^2\CT}{\partial X^{I}\partial X^{J}} - c^{I} \frac{\partial \mathcal{G}_0}{\partial X^{I}} + \frac{c^{0}}{X^0}\Bigl(\frac{\chi}{24}-1\Bigr),\\
&\Upsilon_2= \frac{1}{3} c^{I} c^{J} c^{K} c^{L} \frac{\partial^4 \mathcal{G}_{-2}}{\partial X^{I}\partial X^J\partial X^K\partial X^L} - \frac{2}{3} c^{I} c^J c^K \frac{\partial^3 \CT }{\partial X^{I}\partial X^J\partial X^K}+\nonumber\\
&\hphantom{\Upsilon_2=}{}+ c^{I}c^{J} \frac{\partial^2 \mathcal{G}_0}{\partial X^{I}\partial X^{J}} + \frac{\bigl(c^{0}\bigr)^2}{(X^0)^2}\Bigl(\frac{\chi}{24}-1\Bigr),
\end{align}
where we have denoted the second derivative of the modified prepotential as
\begin{equation}
\tau_{IJ} = {\partial^2\widetilde\CG_{-2} \over \partial X^{I}\partial X^{J}}.
\end{equation}

So far, we have just considered one-instanton solutions satisfying the boundary condition \smash{$\CG_\CA^{(1)}=1$}. We will now consider multi-instanton solutions with more general boundary conditions.

 \subsection{Multi-instantons} \label{sec-multinsts}

In analogy to the closed topological string, following~\cite{gkkm,gm-multi}, we can derive HAEs for the
partition function of the real topological string, and use them to compute multi-instanton contributions. For this purpose, we introduce
\be
\label{zhalf}
Z=\exp\biggl[\frac{1}{2}\mathcal{G}\biggr],\qquad
Z^{(0)}=\exp\biggl[\frac{1}{2}\mathcal{G}^{(0)}\biggr],\qquad Z_{\text{r}}=\frac{Z}{Z^{(0)}}.
\ee
We will call $Z_{\rm r}$ the reduced partition function.
Notice the additional factor of $1/2$ in comparison to the closed string case discussed in~\cite{gkkm,gm-multi}, which is necessary in order to find linear equations for the reduced partition function. These equations read
\begin{align}
\begin{aligned}
& \mathsf{W}Z_{\text{r}} = \mathsf{D}^2 Z_{\text{r}},\\
& \mathsf{R}Z_{\text{r}} = - \mathsf{D} Z_{\text{r}}.
\end{aligned}\label{zr-eq}
\end{align}
We use a trans-series ansatz for $Z_{\rm r}$ with two trans-series parameters $\CC_1$ and $\CC_2$,
as in~\cite{gkkm,gm-multi},
\begin{equation}
Z_{\text{r}} = 1+\sum\limits_{(n,m)\neq(0,0)} \CC_1^n\CC_2^m Z_{\text{r}}^{(n|m)}.
\end{equation}
This generalizes our previous ansatz (\ref{one-par-ts}). The sector $(n|m)$ corresponds to $n$ instantons and $m$ ``negative'' instantons,
and it behaves at small $g_{s}$ as
\begin{equation}
\label{eq:multi-inst-small-gs}
Z_{\text{r}}^{(n|m)}\sim\exp\biggl(-(n-m)\frac{\mathcal{A}}{g_{s}}\biggr).
\end{equation}
These mixed sectors were first considered in a related context in~\cite{gikm}, see~\cite{asv11,mss} for further developments.
From (\ref{zr-eq}), we find, by linearity,
\begin{align}
\begin{aligned}
&\mathsf{W}Z_{\text{r}}^{(n|m)}= \mathsf{D}^2 Z_{\text{r}}^{(n|m)},\\
& \mathsf{R}Z_{\text{r}}^{(n|m)}= - \mathsf{D} Z_{\text{r}}^{(n|m)}.
\end{aligned}
\end{align}
These equations can now be treated with the previously introduced operator formalism. We make the ansatz
\begin{equation}
\label{eq:anstatz-multi}
Z_{\text{r}}^{(n|m)} = \mathfrak{a}_{(n|m)}\text{e}^{\Sigma^{(n|m)}}.
\end{equation}
Then one finds that $\Sigma^{(n|m)}$ satisfies the following equations:
\begin{align}
\label{eq:haephi1compact}
&\mathsf{W}\Sigma^{(n|m)}= \mathsf{D}^2 \Sigma^{(n|m)} + \bigl(\mathsf{D}\Sigma^{(n|m)}\bigr)^2,\\
\label{eq:haephi2compact}
&\mathsf{R}\Sigma^{(n|m)}= - \mathsf{D} \Sigma^{(n|m)},
\end{align}
while $\mathfrak{a}^{(n|m)}$ satisfies
\begin{align}
\begin{aligned}
&\mathsf{W}\mathfrak{a}^{(n|m)} = \mathsf{D}^2 \mathfrak{a}^{(n|m)} + 2 \bigl(\mathsf{D}\Sigma^{(n|m)}\bigr)\bigl(\mathsf{D}\mathfrak{a}^{(n|m)}\bigr), \\
&\mathsf{R}\mathfrak{a}^{(n|m)} = -\mathsf{D} \mathfrak{a}^{(n|m)}.
\end{aligned}
\end{align}
The equations above only differ in numerical prefactors from their closed string siblings given in~\cite{gkkm,gm-multi}. Let us proceed to construct solutions to the above equations starting with $\Sigma^{(n|m)}$. Consider first with the object $H$ introduced in~\eqref{eq:def-H}. We claim that, in complete analogy to the construction in Section~\ref{subsec:one-inst-amp-op}, we have
\begin{equation}
\Sigma^{(n|m)} = \mO^{(n-m)}H,
\ee
where we have introduced the operator
\be
\mO^{(\ell)}=\sum\limits_{k=1}^{\infty}\frac{2^{k-1}}{k!}(-\ell)^{k}\mathsf{D}^{k-1} = \frac{1}{2\mathsf{D}}\bigl(\text{e}^{-2\ell\,\mathsf{D}}-1\bigr) = -\frac{1}{2}\int_{0}^{2\ell} \text{e}^{-u\mathsf{D}}\text{d}u.
\end{equation}
Indeed, for the first equation \eqref{eq:haephi1compact} the proof is identical to the one for the one-instanton amplitude, whereas for the second equation \eqref{eq:haephi2compact} we remember that
\begin{equation}
[\mathsf{R}, \mathsf{D}] = 0,
\end{equation}
which means that \smash{$\bigl[\mathsf{R}, \mO^{(\ell)}\bigr] = 0$}. We conclude that equation \eqref{eq:haephi2compact} follows directly from equation~\eqref{J-eqns}.

We are left to investigate the prefactor $\mathfrak{a}^{(n|m)}$. To this end, it will be useful to introduce the operator
\begin{equation}
\mathsf{M}^{(n|m)} = \mathsf{W} - \mathsf{D}^2 - 2 \bigl(\mathsf{D}\Sigma^{(n|m)}\bigr)\mathsf{D}.\\
\end{equation}
Then we want to solve\footnote{Notice that $\mathsf{M}^{(n|m)} $ differs from its counterpart in~\cite{gkkm} only by factors of 2.}
\begin{align}
\label{eq:cond-pref-multi-1}
&\mathsf{M}^{(n|m)}\mathfrak{a}^{(n|m)}= 0,\\
\label{eq:cond-pref-multi-2}
&\mathsf{R}\mathfrak{a}^{(n|m)}= - \mathsf{D} \mathfrak{a}^{(n|m)},
\end{align}
subject to the boundary condition
\begin{equation}
\label{eq:bound-cond-multi}
\mathfrak{a}_{(n|m), \mathcal{A}} = \biggl(\frac{\mathcal{A}}{g_{s}}\biggr)^{\ell}.
\end{equation}
As already explained in~\cite{gkkm,gm-multi}, by linearity of \eqref{eq:cond-pref-multi-1} and \eqref{eq:cond-pref-multi-2},
this suffices to obtain solutions whose boundary conditions are general polynomials in $\mathcal{A}/g_s$. We
will now construct objects $\mathfrak{m}_{\ell}$, $\ell=1,2, \dots$,
that fulfill equations \eqref{eq:cond-pref-multi-1} and \eqref{eq:cond-pref-multi-2}, subject to
the boundary condition \eqref{eq:bound-cond-multi}. In complete analogy to the formalism
developed in~\cite{gkkm,gm-multi}, we introduce
\begin{equation}
\label{Xdef}
X=H + 2\,\mathsf{D}\Sigma_{(n|m)},
\end{equation}
which fulfills
\begin{equation}
\mathsf{M}^{(n|m)}X=0.
\end{equation}
(The object $X$ appearing in (\ref{Xdef}) should not be confused with the flat coordinates $X^I$ introduced in (\ref{omperiods}).)
Then we define $\mathfrak{m}_{\ell}$ via
\begin{equation}
\Xi(\xi) = \exp (\mathcal{L}_{\xi}X ) = \sum\limits_{\ell=0}^{\infty} \frac{\mathfrak{m}_{\ell}}{\ell!} \xi^{\ell},
\end{equation}
where
\begin{equation}
\mathcal{L}_{\xi} = \frac{1}{2}\int_{0}^{2\xi}\,\text{e}^{u\mathsf{D}}\text{d}u = \sum_{k \ge 1} {\xi^k \over k!} ( 2 \oD )^{k-1}.
\end{equation}
First we notice that $\Xi(\xi)$ fulfills equation \eqref{eq:cond-pref-multi-2}, by using formulae \eqref{J-eqns},
\eqref{eq:haephi2compact} and the fact that $\mathsf{R}$ commutes with $\mathsf{D}$.
Concerning the first condition \eqref{eq:cond-pref-multi-1}, it is sufficient to show that
\begin{equation}
\mathsf{M}^{(n|m)}\Xi(\xi) =0.
\end{equation}
This can be proven by following exactly the procedure in~\cite{gkkm,gm-multi} and we will not repeat the explicit
steps here. To illustrate the considerations above, we list the first few examples of $\mathfrak{m}_{\ell}$:%
\begin{align}
\begin{aligned}
&\mathfrak{m}_{1}= X, \\
&\mathfrak{m}_{2}= X^2 + 2\mathsf{D}X, \\
&\mathfrak{m}_{3}= X^3 + 6 X \mathsf{D}X + 4 \mathsf{D}^2X.
\end{aligned}
\end{align}
Finally, let us consider the holomorphic limit of the above results. The multi-instanton amplitude is given by formula \eqref{eq:anstatz-multi} and in close analogy to the discussion in Section~\ref{subsec:one-inst-amp-op} we find for the exponent $\Sigma^{(n|m)}$
\begin{equation}
{1\over 2} \bigl(\tilde \CG\bigl(X^I-2(n-m) g_s c^I\bigr)-\tilde \CG\bigl(X^I\bigr) \bigr).
\end{equation}
Since the prefactors appearing in the multi-instanton amplitudes consist of words made out of the letters $\mathsf{D}^k X$, it suffices to establish that
\begin{equation}
\mathsf{D}^{k-1} X \to g_{s}^k c^{I_1}\cdots c^{I_k}\frac{\partial^k}{\partial X^{I_1}\cdots\partial X^{I_k}}\tilde{\mathcal{G}}\bigl(X^{I}-2(n-m)g_{s}\bigr),
\end{equation}
which follows immediately from (\ref{hol-D}) and the holomorphic limits of $H$ and $\Sigma^{(n|m)}$.

\subsection{Boundary conditions}
 \label{sec-bc}
 Boundary conditions play a double role in the HAE approach to trans-series. First, they allow to fix
 the holomorphic ambiguity. In addition, they indicate which type of instanton sectors (and instanton actions) appear in the non-perturbative sectors. To obtain boundary conditions for the trans-series, we follow~\cite{cesv2,cesv1,cms, gkkm} and we consider the
behavior of the theory near special points, like the conifold point and the large radius point. This behavior can be immediately translated
in terms of a large order behavior for the amplitudes at large genus, which leads in turn to multi-instanton amplitudes.
For simplicity, we will assume that we are in a situation with a single modulus.

Let us first consider the conifold point. In the case of the conventional closed topological string, we have the behavior (\ref{gap}).
The formula for the Bernoulli numbers
 \be
 \label{lo-bernoulli}
 B_{2g} = (-1)^{g-1} {2 (2g)! \over (2 \pi)^{2g}} \sum_{\ell \ge 1}\ell^{-2g}
 \ee
 gives the all-orders asymptotic behavior for the pole term in (\ref{gap}):
 \be
 \label{lobern}
 {1 \over 2 \pi^2} \Gamma(2g-1) \sum_{\ell \ge 1} (\ell \CA_c )^{1-2g}\bigl( \mu_{0,\ell} + {\ell \CA_c \over 2g-2} \mu_{1,\ell}\bigr),
 \ee
 where
 \be
 \label{ac-gen}
 \CA_c= 2 \pi \ri t_c, \qquad \mu_{0,\ell} = {\CA_c \over \ell}, \qquad \mu_{1, \ell} ={1\over \ell^2}.
 \ee
The instanton action appearing here is the central charge of the D-brane whose mass vanishes at the conifold, in an appropriate normalization. According to the correspondence between large order behavior and exponentially small corrections (see, e.g.,~\cite{mmbook}),
(\ref{lobern}) corresponds to an~$\ell$-th instanton amplitude of the Pasquetti--Schiappa form~\cite{ps09},
 \be
 \label{psform}
 {1\over 2 \pi} \biggl( {1\over \ell} {\CA_c \over g_s} + {1\over \ell^2} \biggr) \re^{-\ell \CA_c /g_s}.
 \ee
We recall that, since the original series is even in $g_s$, we will have a trans-series with action~$-\CA_c$ as well. They both combine to
give the asymptotic behavior (\ref{lobern}). In general, in the discussion below we have to consider as well the trans-series with opposite action.\footnote{The asymptotic series for the real topological string free energy is obviously not even in $g_s$. However, the boundary conditions associated to the conifold point involve both actions $\pm \CA$.}

In the case of the real topological string, we have an additional term in $\CG_\chi$ when $\chi=2g-2$ is even, given by (\ref{k-gap}) and (\ref{gap-real}). As in the closed string case, we can extract from these terms a large genus behavior. For the first term in (\ref{gap-real}),
we can use (\ref{lo-bernoulli}), and we find
 \be
 - {t_c^{2-2g} \over 2^{2g+1} g(g-1)} \bigl(2^{2g}-1\bigr) B_{2g}= -{1\over \pi^2} \Gamma(2g-1)\sum_{\ell \, \text{odd}} (\ell \CA_c)^{2-2g} {1\over \ell^2} \biggl( 1+ {1\over 2g-2} \biggr).
 \ee
The second term in (\ref{gap-real}) gives
 \be
 {t_c^{2-2g} \over 2^{2g+1} (g-1)} E_{2g-2}=
 {1\over \pi} \Gamma(2g-2) \sum_{\ell \, \text{odd}} ( \ell \CA^{\rm o}_c)^{2-2g} {(-1)^{\ell-1 \over 2} \over \ell},
 \ee
 where
 \be
 \label{aco}
 \CA_c^{\rm o}= {1\over 2} \CA_c.
 \ee
 The first term leads to a multi-instanton correction which is minus twice the Pasquetti--Schiappa form (\ref{psform}),
 \be
-{1 \over \pi } \biggl( {1\over \ell} {\CA_c \over g_s} + {1 \over \ell^2} \biggr) \re^{-\ell \CA_c/g_s} ,
 \ee
 and in addition $\ell$ takes only positive, odd integer values. The second term corresponds to a~multi-instanton correction of the form
 \be
 \label{leading-mi}
 {(-1)^{\ell-1 \over 2} \over \ell}\re^{-\ell \CA_c^{\rm o}/g_s} ,
 \ee
 where $\ell$ is also odd. The appearance of half the instanton action (\ref{aco}) is probably an effect of the orientifold action. Note that
 the leading multi-instanton corresponds to (\ref{leading-mi}) with $\ell=1$.

We conclude that the instanton actions or Borel singularities include the sequence
\be
\label{mhalf-action}
 (2k +1 ) \CA, \qquad k \in \IZ_{\ge 0},
\ee
where $\CA =\CA_c^{\rm o}$. They lead to the boundary condition
\be\label{bc-aco}
\CG_{\CA}^{(2k+1)}= {(-1)^k \over 2k+1}\re^{-(2k+1) \CA/g_s}, \qquad k=0,1,2,\dots\,.
\ee
This is what we used in (\ref{bc-ex}), with $k=0$. We also have the sequence of Borel singularities
\be
2 k \CA, \qquad k \in \IZ_{>0}.
\ee
For $k$ even, they are due to the contributions from the closed string sector. When $k$ is odd,
we have contributions from both the closed and the non-orientable sector. The resulting boundary condition is
\be\label{ps-con}
\CG_{ \CA}^{(2 k)}={(-1)^k \over 2\pi } \biggl( {2\over k} {\CA \over g_s} + {1 \over k^2} \biggr)\re^{-2k \CA /g_s}, \qquad k=1,2, \dots\, .
\ee
This analysis shows that the minimal instanton action is {\it half} the D-brane central charge, as~shown in (\ref{aco}). It is natural to assume that this is due to the orientifold action, but a~further understanding of this point would require a space-time theory of the topological string instantons. In any case, this suggests that the relevant actions in the theory are half-integral periods of the holomorphic 3-form, as anticipated above.

Let us now consider the large radius point. It was shown in~\cite{gkkm}, in the closed string case, that the GV formula (\ref{gv-F}) determines
the large genus asymptotics near the large radius point, and one can read from it
the location of a sequence of instanton actions and the corresponding multi-instanton amplitudes and Stokes constants. It turns out that
this asymptotics is determined by the genus zero GV invariants. Indeed, by expanding (\ref{gv-F}) in powers of $g_s$, we find
\be
\label{fg-gv}
\CF_g(t)= \sum_{d\ge 1} \biggl( {(-1)^{g-1} B_{2g} \over 2g (2g-2)!} n_{0,d}+ {2 (-1)^g n_{2, d} \over (2g-2)!}+ \cdots \biggr) {\rm Li}_{3-2g}\bigl(\re^{-d\cdot t} \bigr).
\ee
Only the first term inside the parentheses in the right-hand side of (\ref{fg-gv}) leads to factorial growth. By
using again the asymptotics (\ref{lo-bernoulli}) and the identity
\be
\label{polylog}
\sum_{n \in \IZ} {1\over (2 \pi \ri n + t)^{m}}= {1 \over (m-1)!} {\rm Li}_{-m+1} \bigl(\re^{-t}\bigr), \qquad m \ge 2,
\ee
one finds that the contribution of genus zero GV invariants leads to instanton actions of the~form%
\be
\label{adm}
\CA_{d, m}=2 \pi d \cdot t+ 4 \pi^2 \ri m , \qquad m \in \IZ.
\ee
The corresponding $\ell$-instanton amplitudes are again of the Pasquetti--Schiappa form
\be
\label{closed-ps}
\CF^{(\ell)}_{\CA_{d,m}} = {n_{0,d} \over 2 \pi} \biggl( {1\over \ell} {\CA_{d,m} \over g_s} + {1 \over \ell^2} \biggr) \re^{-\ell \CA_{d,m}/g_s},
\ee
and $n_{0,d}$ is the corresponding Stokes constant (up to normalization).

In the case of the real topological string, one has to consider the additional contribution (\ref{BPS-real}). The only source of
factorial growth is due to the term with $r=-1$, and one has
\begin{gather}
\CG_{\chi=2s-1}\!= \tilde n_{-1, d} {\ri(-1)^s B_{2s} \bigl(2^{2s-1}-1\bigr) \over (2s)!} \bigl\{ 2^{2-2s} {\rm Li}_{2-2s} \bigl(\re^{-d\cdot t/2}\bigr)
-{\rm Li}_{2-2s} \bigl(\re^{-d\cdot t}\bigr)\! \bigr\}+ \cdots,\!\!
\end{gather}
where the degrees $d$ are odd positive numbers. This leads to a large $\chi$ asymptotics which can be
decoded in terms of multi-instanton amplitudes of the form
\be\label{new-lr}
\ri \tilde n_{-1, d}c_{\CA} { (-1)^{\ell-1} \over \ell} \re^{-\ell \CA/g_s}, \qquad \ell \in \IZ_{>0},
\ee
where $\CA$ can take the following values:
\be
\label{possible-as}
{1\over 2} \CA_{d, m}, \qquad {1\over 2} \CA_{d, 2m}.
\ee
Here, $\CA_{d,m}$ is given in (\ref{adm}), and the coefficient $c_\CA$ in (\ref{new-lr}) takes the values $1$ and
$-2$ for the actions given in (\ref{possible-as}), respectively. Note that (\ref{possible-as}) requires again half-integral periods, and that, when $\ell$ is even, the multi-instanton amplitude (\ref{new-lr}) will combine with multi-instanton amplitudes (\ref{closed-ps}) associated to the closed string sector.

This analysis leads to two main conclusions. First, the real topological string leads to a~new type
of boundary condition, encoded in the new multi-instanton amplitudes of the form~(\ref{leading-mi}) and~(\ref{new-lr}), and involving half-integral periods.
Second, in the resurgent structure
of the real topological string, the disk invariants $\tilde n_{-1,d}$
appear as Stokes constants associated to the new singularities (\ref{possible-as}).

\subsection{Stokes automorphisms}

The analysis in the previous section indicates that there are two new types of trans-series sectors
appearing in the resurgent structure of this theory.
The first one is associated to the new boundary condition at the conifold (\ref{bc-aco}) and (\ref{ps-con}), while the second one is associated to the new boundary condition at large radius (\ref{new-lr}). A compact way to encode these trans-series sectors is to obtain the corresponding
Stokes automorphisms, as it was done in~\cite{im} for the closed topological string.

Let us consider the trans-series associated to the conifold behavior. They are multi-instanton amplitudes corresponding to the
action $\CA= \CA_c^{\rm o}$. In the $\CA$-frame, the $\ell$-instanton amplitudes with $\ell$ odd are given by (\ref{bc-aco}), while the ones with $\ell$ even are given by (\ref{ps-con}). We first calculate the generating function
\be
\label{gen-stokes}
\sum_{\ell \ge 1}\CC^{\ell} \CG_{\CA}^{(\ell)}= {\CB}_c\biggl(\CC\re^{-\CA/g_s}, {\CA \over g_s} \biggr),
\ee
where
\be
\CB_c (x,y)= \tan^{-1} (x)+{1\over 2 \pi} {\rm Li}_2 \bigl(-x^2\bigr)-{y \over \pi} \log\bigl(1+ x^2\bigr).
\ee
The Stokes automorphism is defined by (see, e.g.,~\cite{abs, msauzin} for background on alien derivatives and Stokes automorphism)
\be
\mathfrak{S}_\CC = \exp \Biggl( \sum_{\ell =1}^\infty \CC^{\ell} \dot \Delta_{\ell \CA} \Biggr).
\ee
Let us now define the partition function $\widetilde Z$ as in (\ref{zhalf}),
\be\label{tildez}
\widetilde Z= \re^{\tilde \CG/2},
\ee
and let us write the instanton action
as a linear combination of periods, as in (\ref{action-periods}). Then, just like in~\cite{im}, there are two different situations.
If all the $c^I$ vanish, the Stokes automorphism acts simply as a global multiplication factor,
\be
\mathfrak{S}_\CC \bigl(\widetilde Z\bigr)= \exp \biggl[ {a \over 2} \CB_c \biggl( \CC \re^{-\CA/g_s}, {\CA \over g_s} \biggr) \biggr] \widetilde Z,
\ee
where $a$ is a Stokes constant (for the singularities at integer multiples of
$\CA= \CA_c^{\rm o}$, one has $a=1$, but one could have
singularities with non-trivial Stokes constants). The factor of $1/2$ in the exponent is
inherited from the definition of \smash{$\widetilde Z$} in
(\ref{tildez}).
When not all $c^I$ vanish, an argument similar to the one in~\cite{im} gives the following formula:
\be\label{stokes-coni}
\mathfrak{S}_\CC (\widetilde Z)= \exp \biggl[ { a \over 2} \CB_c \bigl( \CC \re^{-2\kappa g_s c^I \partial_I}, 2\kappa g_s c^I \partial_I \bigr) \biggr] \widetilde Z,
\ee
where we have re-introduced the normalization factor $\kappa$ defined in (\ref{action-periods}).
The expression (\ref{stokes-coni}) for the Stokes automorphism is rather
different from the one obtained in~\cite{im}, and from similar transformations that
have appeared in the literature (see, e.g.,~\cite{AP, teschner2}). However, as in~\cite{im}, it is essentially determined by
the conifold behavior of the free energies.

The Stokes automorphism associated to the multi-instanton amplitude (\ref{new-lr}) can be worked out in a similar way,
and it involves the simpler generating function
\be
\CB_{\rm LR}(x)=\log( 1+x).
\ee
As we noted above, this contribution has to be combined with the one due to the closed string sector, but the resulting transformation
can be found in a straightforward way.

\section[Experimental evidence: the real topological string on local P\^{}2]{Experimental evidence:\\the real topological string on local $\boldsymbol{\IP^2}$}
\label{sec-p2}
In this section, we present experimental evidence for our non-perturbative results, based on the connection between
instanton amplitudes and large order behavior of the perturbative series.

\subsection{Perturbative expansion}
%\label{subsec:pert-exp}

Our basic example in this section is the real topological string on local $\IP^2$ mentioned in Section~\ref{review-TS}, where the choice of involution is given by complex conjugation on both the fiber and the base. This model was studied in detail in
\cite{kw-localp2, walcher-tadpole}, which we will follow closely. Of course, the closed string sector is well known, and the closed string topological free energy can be computed systematically with the conventional HAE, see, e.g.,~\cite{cesv2, hkr}. Some aspects of the special geometry of the model are summarized in Appendix~\ref{app-localp2}.

The first new ingredient in the real topological string is the disk amplitude or domain wall tension. At large radius, it is given by
\begin{align}
\CT(z)&{}=\xi 2\Gamma^2(3/2)\sum_{m \ge 0} {\Gamma \bigl(3 m+\frac{3}{2}\bigr) \over \Gamma
 \bigl(m+\frac{3}{2}\bigr)^3}(-1)^m z^{m+\frac{1}{2}}\nonumber\\
 &{}= 2 \xi {\sqrt{z}}\,
 _4F_3\biggl(\frac{1}{2},\frac{5}{6},1,\frac{7}{6};\frac{3}{2},\frac{3}{2},\frac{3}{2};-
 27 z\biggr).\label{TLR}
 \end{align}
Here, $\xi$ is a normalization factor which has to be chosen appropriately to have a
 coherent addition of the different sectors of the real topological string. We will usually set
 \be
 \xi= \ri.
 \ee
The complex variable $z$ parametrizes the moduli space of complex structures of local $\IP^2$. The conifold point occurs at $z=-1/27$, while $z=0$ is the large radius point. Note also that the sign of $z$ is the opposite one to what is
 used in~\cite{kw-localp2,walcher-tadpole}. We can now use the integrality formula following from (\ref{BPS-real})
 \be
 {1\over \xi} \CT= 2 \sum_{k,d \ \text{odd}} {1\over k^2} \tilde n_{-1,d} Q^{dk/2},
 \ee
 where $Q=\re^{-t}$, to obtain the counting of disks
 \be
 \sum_{d \ \text{odd}} \tilde n_{-1,d} Q^{d/2}= Q^{1/2}- Q^{3/2}+ 5 Q^{5/2} -42 Q^{7/2}+429 Q^{9/2}+ \cdots.
 \ee
One interesting observation in~\cite{walcher-tadpole} is that the domain wall tension
(\ref{TLR}) can be obtained as the difference between the disk amplitudes of~\cite{akv, av}, evaluated at two different
values of the open moduli. We give some details of this computation in Appendix~\ref{app-disk-av}.

We will be interested in evaluating the real topological string amplitudes in different frames. A~particular important one is the conifold frame,
in view of the behavior (\ref{gap}),~(\ref{k-gap}) and~(\ref{gap-real}). The appropriate flat coordinate in this frame is
denoted by $t_c(z)$ and defined in (\ref{tc}). It vanishes at the conifold point of the local $\IP^2$ geometry, and it has an expansion
as a power series in the local coordinate around the conifold $\delta=1+27 z$, defined in (\ref{con-lc}). The domain wall tension~$\CT$ in this frame can be simply
 obtained by noting that $\CT$ solves the inhomogeneous Picard--Fuchs equation
 \be
{1\over \xi} \mathfrak{L} \CT= -{\sqrt{-z}\over 4}.
\ee
According to the discussion around (\ref{hol-grif}), the domain wall tension in the conifold frame should vanish quadratically at the conifold
point,
 \be
 \CT_c = a \delta^2+ \CO\bigl(\delta^3\bigr).
 \ee
 One finds
 \be
 \label{Tc}
\CT_c= \frac{\delta ^2}{24 \sqrt{3}}+\frac{121 \delta ^3}{2592 \sqrt{3}}+\frac{3197 \delta
 ^4}{69984 \sqrt{3}}+\frac{4372889 \delta ^5}{100776960 \sqrt{3}}+\CO\bigl(\delta
 ^6\bigr).
 \ee
As expected from the discussion in (\ref{Tgamma}), $\CT_c$ does not agree
with the large radius tension (\ref{TLR}). Let us take the closed string periods to be
 given by $t_\IR (z)$, $t_c(z)$ and $1$, where
 \be
 t_\IR= -\log(-z) - \widetilde \varpi_1(z),
 \ee
and $\varpi_1 (z)$ is the series in (\ref{varpisp2}). Then, one finds
 \be
 \CT_c- \CT= \alpha_0 +\alpha_1 t_\IR (z) + \beta t_c(z),
 \ee
 where the coefficients $\alpha_{0,1}$ and $\beta$ are given by
 \be
 \label{coeffs-t}
 \alpha_0= V, \qquad \alpha_1= -{\pi \over 6}, \qquad \beta= {\log(2) \over {\sqrt{3}}},
 \ee
 and $V$ is given by
\be
\label{Vdef}
V= 2 \operatorname{Im} {\rm Li}_2 \bigl(\re^{\pi \ri/3}\bigr).
\ee
These numbers were computed numerically to very high precision
 and then we fitted them to conjectural exact expressions by using number recognition in WolframAlpha.
 By using the value of $t_\IR(z)$ at the conifold point (see, e.g.,~\cite{dk,mz, rv}),
 \be
 t_\IR \biggl( -{1\over 27} \biggr)= {9 V \over 2 \pi},
 \ee
 we obtain the conjecture
 \be
 \label{con-tension}
 \CT \biggl(-{1\over 27} \biggr)= -{V\over 4}.
 \ee
 It would be very interesting prove the conjectures for the coefficients (\ref{coeffs-t}) and for the value of~$\CT$ at the conifold (\ref{con-tension})
 by extending the techniques and ideas of~\cite{bksz,dk, rv}.

 Once the domain wall tension is known, we can calculate the holomorphic limits
 of the Griffiths invariant and of the propagator $\CR^z$ for local $\IP^2$ (remember that \smash{$\tilde R$} can be set to
 zero in the local case). For the Griffiths invariant
 we can use (\ref{hol-grif}). If in addition we use flat coordinates, the covariant
 derivatives become conventional derivatives, and one has
 \be
 {\cal D}_{tt}= \partial_t^2 \CT.
 \ee
 The holomorphic limit of the propagator follows then from (\ref{hol-nice}), where one chooses $d_{zz}=0$~\cite{kw-localp2},%
 \be
 \CR^z=- {1\over C_z} {\cal D}_{tt} \bigl(t'(z) \bigr)^2.
 \ee
 Here, as in~\cite{gm-multi}, we have denoted by $C_z$ the only entry of the Yukawa coupling $C_{ijk}$
 in the one-modulus case (its explicit expression in the
 case of local $\IP^2$ can be found in (\ref{yc-p2})). We recall from~\cite{gm-multi} that the holomorphic limit of the closed string propagator in the local, one-modulus case can be written as
 \be
 \label{S-onemod}
 \CS= -{1\over C_z} {t''(z) \over t'(z)} - \mathfrak{s},
 \ee
 where $\mathfrak{s}= -s^{z}_{zz}/C_z$ and we have denoted $\CS\equiv \CS^{zz}$. It follows from this expression and~(\ref{hol-limit}), (\ref{rk-cons}) that the real propagator
 can be also written as
 \be
 \CR^z={1\over C_z} \biggl( {t''(z) \over t'(z)} {\partial \CT \over \partial z} - {\partial^2 \CT \over \partial z^2 } \biggr),
 \ee
 since $h_{zz}=d_{zz}=0$. This is the real counterpart of the closed string formula (\ref{S-onemod}), and it is valid for arbitrary
 frames. In the conifold frame, for example, one finds
 \be
\CR_c^z= -\frac{t_c}{108}+\frac{53 t_c^2}{1296\sqrt{3}}-\frac{817 t_c^3}{23328}+\frac{346487 t_c^4}{5038848 \sqrt{3}}+\CO\bigl(t_c^5\bigr).
 \ee
This agrees with the result in~\cite{kw-localp2} up to a rescaling\footnote{Notice how our conventions in~\cite{gm-multi} differ from ours by the rescaling $t_c\to t_c/\sqrt{3}$.} of $t_c$ by $3$.

With these ingredients, one can calculate the higher $G_\chi$ by using (\ref{djg0}) for $\chi=0$ and Walcher's HAE
for higher values of $\chi$. The only non-trivial ingredient is the fixing of the holomorphic ambiguity
at each value of $\chi$. As in~\cite{kw-localp2}, we do this by combining the conifold behavior~(\ref{gap}) and~(\ref{gap-real}) with an
explicit calculation of the real topological string free energy with the real topological vertex
of~\cite{ksw, kw-localp2}. This
calculation is time-consuming and sets a~practical limit to the number of terms
that we can compute. We have obtained explicit results
up to~${\chi=22}$. This is not such a long perturbative series, as compared, e.g., to what
was used in~\cite{gm-multi} for the closed string free energies, but it is enough to check
the asymptotic predictions, as we will see in the next section.

\subsection{Trans-series and asymptotics}

We will now test the formulae derived in Section~\ref{sec:trans-series-res-structure} for the case of
the real topological string on local $\mathbb{P}^2$, by using the resurgent connection between instanton amplitudes
and large order behavior of the perturbative series (see, e.g.,~\cite{mmbook}). Our series will be given by the perturbative real string
free energy in the large radius frame, and for simplicity we will focus on the region of moduli space where $-1/27<z<0$.

The Borel singularity that controls the asymptotics of the perturbative sector not too far from the conifold point
is set by the conifold behavior (\ref{gap})--(\ref{gap-real}). As we found in Section~\ref{sec-bc}, the smallest action associated to this behavior is given by
\begin{equation}
\label{eq:prop-action}
\CA=\mathcal{A}_c^{\rm o} = \pi \ri t_c,
\end{equation}
where $\mathcal{A}_c^{\rm o}$ was defined in (\ref{aco}), The corresponding trans-series is
determined by the boundary condition (\ref{leading-mi}) with $\ell=1$, or \eqref{bc-ex}, which fixes
as well the overall coefficient or Stokes constant. It is given by the general expression \eqref{eq:One-Inst-HAE-Sol}, which we repeat here for the convenience of the reader,
\be
\label{eq:one-instanton-local-one-mod}
\CG^{(1)}= \exp\biggl[{1\over 2} \bigl(\tilde \CG(t-2c g_s )-\tilde \CG(t)\bigr) \biggr].
\ee
In this equation, the coefficient $c$ is given by
\begin{equation}
c=\frac{3 \ri}{2},
\end{equation}
as it follows from \eqref{eq:prop-action} and the results in Appendix
\ref{app-localp2} ($c$ is half the constant $\alpha$ in~\cite{gm-multi}).

We are now ready to perform the explicit asymptotic checks of the perturbative series. It is more convenient to consider a real action and therefore in what follows we will rescale the real free energy as
\begin{equation}
\mathcal{G}_{\chi}\,\to\,(-\ri)^{\chi}\mathcal{G}_{\chi}, \qquad\mathcal{A}\to-\ri\mathcal{A}.\nonumber
\end{equation}
From our analytical prediction, and taking into account the fact that instanton actions appear in pairs $\CA$, $-\CA$, we find the large order formula
\begin{equation}
\label{eq:one-inst-contribution}
\CG_\chi \sim \frac{1}{\pi} \mathcal{A}^{-\chi} \Gamma(\chi) \biggl( \mu_0 + {\mu_1 \CA \over \chi}+ \cdots \biggr), \qquad \chi \gg 1.
\ee
In this equation, $\mu_0$ is given by
\be
\label{mu0-thv}
\mu_0= \exp\bigl(c^2\partial_t^2\tilde{\mathcal{G}}_{-2}(t)\bigr)
\begin{cases}
\cosh\bigl(c\partial_{t}\tilde{\mathcal{G}}_{-1}(t)\bigr) &\text{if }\chi\text{ even},\\
\sinh\bigl(c\partial_{t}\tilde{\mathcal{G}}_{-1}(t)\bigr) &\text{if }\chi\text{ odd},
\end{cases}
\end{equation}
while $\mu_1$ is
\be
\label{eq:nloasymptotics}
\mu_1=\exp\bigl(c^2\partial_t^2\tilde{\mathcal{G}}_{-2}(t)\bigr)\begin{cases}
\zeta_1\cosh\bigl(c\partial_{t}\tilde{\mathcal{G}}_{-1}(t)\bigr)+\zeta_2\sinh\bigl(c\partial_{t}\tilde{\mathcal{G}}_{-1}(t)\bigr), & \chi\text{ even},\\[5pt]
\zeta_1\sinh\bigl(c\partial_{t}\tilde{\mathcal{G}}_{-1}(t)\bigr)+\zeta_2\cosh\bigl(c\partial_{t}\tilde{\mathcal{G}}_{-1}(t)\bigr), & \chi\text{ odd},
\end{cases}
\ee
where
\be
\zeta_1= c^2\frac{\partial^2 \CT}{\partial t^2},\qquad
\zeta_2=-\frac{2}{3}c^3 C_{ttt} - c \frac{\partial \mathcal{G}_0}{\partial t}.
\ee

\begin{figure}[t]
\centering
\includegraphics[scale=1]{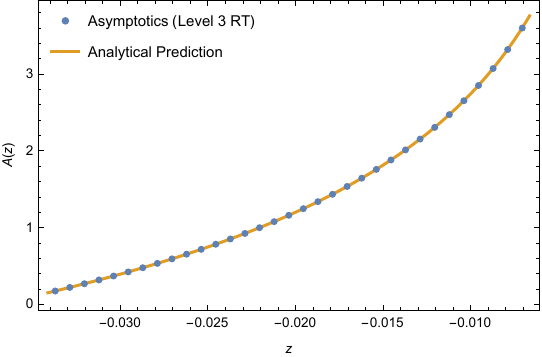}
\vspace{-1mm}

\caption{The continuous line is the expected value for the action $\CA=\pi t_c$, as a function of $z$, while the dots are numerical approximations based on extrapolation and acceleration of the sequence (\ref{action-seq}).}
\label{fig:asymptoticcheckaction}
\end{figure}

\begin{figure}[t]
\centering
\includegraphics[scale=0.75]{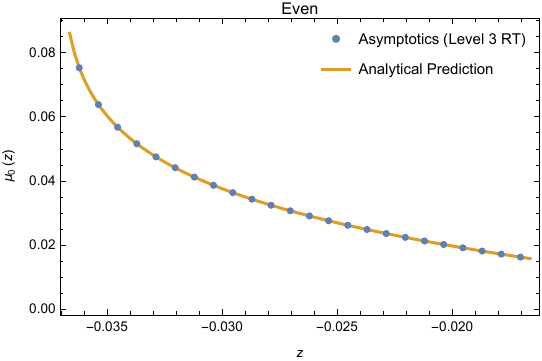}
\qquad
\includegraphics[scale=0.75]{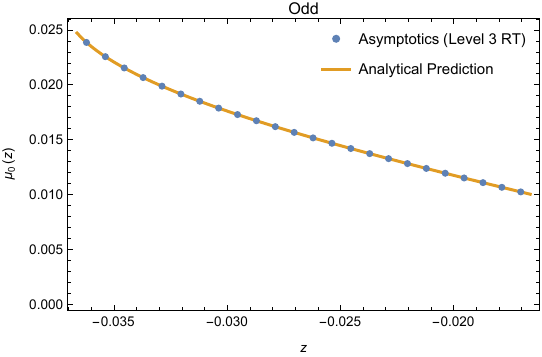}
\vspace{-1mm}

\caption{The continuous line is the expected value (\ref{mu0-thv}) for the coefficient $\mu_0$, as a function of $z$, while the dots are numerical approximations based on extrapolation and acceleration of the appropriate auxiliary sequence. The figure on the left corresponds to even values of $\chi$, while the figure on the right corresponds to odd values. }
\label{fig:asymptoticcheckzeroorder}
\end{figure}

\begin{figure}[!ht]
\centering
\includegraphics[scale=0.75]{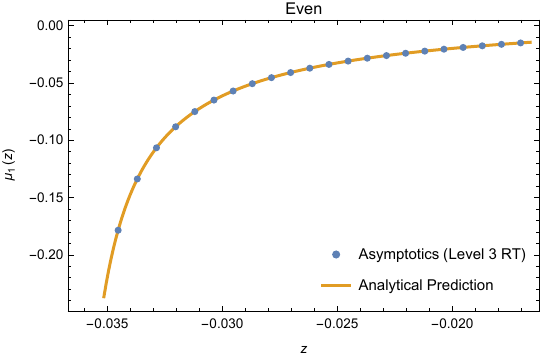}
\qquad
\includegraphics[scale=0.75]{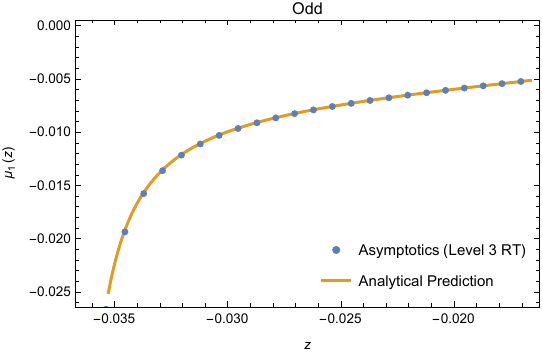}
\vspace{-1mm}

\caption{The continuous line is the expected value (\ref{eq:nloasymptotics}) for the coefficient $\mu_1$, as a function of $z$, while the dots are numerical approximations based on extrapolation and acceleration of the appropriate auxiliary sequence. The figure on the left corresponds to even values of $\chi$, while the figure on the right corresponds to odd values.}
\label{fig:nloasymptoticcheck}
\end{figure}

As usual in resurgence, we can test these predictions by constructing auxiliary
series that converge to the quantities $\CA$, $\mu_{0,1}$ appearing
in the asymptotic formula (\ref{eq:one-inst-contribution}), as in, e.g.,~\cite{msw}. For example,
according to (\ref{eq:one-inst-contribution}) the action $\CA$ should be the limiting value of the sequence
\be
\label{action-seq}
\sqrt{\frac{\mathcal{G}_{\chi}}{\mathcal{G}_{\chi-2}}(\chi-1)(\chi-2)}
\ee
as $\chi \rightarrow \infty$. A numerical approximation to this limit can be obtained by first calculating
the first~$N$ terms in this sequence (in our case, we have
computed them up to $N=22$), and then by using acceleration methods,
like Richardson transforms (RT), to reach higher precision.
 In~Figure~\ref{fig:asymptoticcheckaction}, we compare the numerical values obtained after three RTs, which we represent by points, and
 the theoretical value $\CA=\pi t_c$, which is represented by a continuous line. We find an agreement of 4 to 5 digits depending on the value of $z$.
Similar comparisons can be made for $\mu_0$ and $\mu_1$, which we show in Figures~\ref{fig:asymptoticcheckzeroorder} and~\ref{fig:nloasymptoticcheck}, respectively. In all cases we use $22$ terms of the auxiliary series and three RTs. We find again an agreement
of 4 to 5 digits, which is an excellent one given that the number of terms available is rather small.


\section{Conclusions}\label{sec-conclusions}

In this paper, we have initiated the study of the resurgent structure of Walcher's real topological string. By generalizing
the ideas of~\cite{cesv2, cesv1} and the operator formalism of~\cite{gkkm,gm-multi}, we have obtained multi-instanton amplitudes of the real
topological string as trans-series solutions of Walcher's HAE. The resulting resurgent
structure is slightly more complicated than in the closed string case, since
the boundary conditions coming from the large radius and the conifold behavior lead to
different trans-series. However, as in the closed string case,
the multi-instanton amplitudes are essentially obtained by an integer shift of the closed
string background, while the D-brane and orientifold background is unaffected.

We find it remarkable that the operator formalism of~\cite{gkkm,gm-multi}
can be extended naturally to the real case. The underlying reason might be that, as shown in
\cite{ooguri-real, nw}, one can relate the solutions of Walcher's extended HAE, to solutions of the HAE of~\cite{bcov}
for the closed topological string. Perhaps
the observations of~\cite{ooguri-real, nw} lead to a simpler derivation of the operator formalism obtained in this paper.

The full resurgent structure of the real topological string involves additional Stokes constants,
as compared to the closed string case. It is natural to
conjecture that the new Stokes constants are related to the counting of BPS states in the presence
of D-branes and orientifold planes. Concrete evidence for this connection has been obtained
in Section~\ref{sec-bc}, where we found that the integer invariants counting disks indeed appear as
Stokes constants in the resurgent structure. More generally, our results indicate that the
theory of Donaldson--Thomas invariants underlying BPS counting has a natural extension to the real case.
It would be very interesting to develop these observations
further, and to find an interpretation for the new Stokes automorphism formula (\ref{stokes-coni}) in the
spirit of~\cite{AP, KS}.
Another interesting direction is the study of the real topological string on compact CYs from the point of view of resurgence.
In this endeavour, a~better understanding of the free energies at high Euler characteristic
would be very useful, and it is clearly an interesting problem in itself.

\appendix

\section[Local P\^{}2]{Local $\boldsymbol{\IP^2}$}\label{app-localp2}

\subsection{Useful formulae}
In this appendix, we collect some explicit formulae for the special geometry of local $\IP^2$ which can be found in, e.g.,~\cite{coms,cesv2, hkr}.

The periods solve the Picard--Fuchs equation $\mathfrak{L} \varphi(z)=0$, where
\be\label{p2PF}
\mathfrak{L}=\theta_z^3 + 3z \theta_z (3\theta_z + 1) (3\theta_z + 2)
\ee
and $\theta_z =z \rd /\rd z$. They are built from the power series
\begin{align}
\begin{aligned}
&\widetilde \varpi_1(z)= \sum_{j\ge 1} 3 {(3j-1)! \over (j!)^3} (-z)^j, \\
&\widetilde \varpi_2(z)=\sum_{j \ge 1}{ 18\over j!}
{ \Gamma( 3j )
\over \Gamma (1 + j)^2} \{ \psi(3j) - \psi (j+1) \}(-z)^{j},
\end{aligned}\label{varpisp2}
\end{align}
where $\psi(w)$ is the digamma function. One has
\be
t=-\log(z) -\widetilde \varpi_1(z).
\ee
The prepotential $F_0(t)$ is defined by
\be
\partial_t F_0(t)= {\omega_2 (z) \over 6},
\ee
where
\be
\omega_2(z)= \log^2(z) + 2 \widetilde \varpi_1(z) \log(z) + \widetilde \varpi_2(z).
\ee
The conifold discriminant is
\be\label{con-lc}
\delta= 1+ 27 z.
\ee
The Yukawa coupling is given by
\be\label{yc-p2}
C_z= -{1\over 3 z^3 \delta(z)},
\ee
The flat coordinate at the conifold is
\be\label{tc}
t_c(z)= { \mathfrak{c} \over 6} \bigl(\omega_c(z) -\pi^2 \bigr), \qquad \mathfrak{c}= {3 \over 2 \pi},
\ee
where
\be\label{omc-def}
\omega_c(z)= \log^2(-z)+2 \log(-z) \widetilde \varpi_1(z) + \widetilde \varpi_2(z).
\ee
It has the property that
\be
t_c\biggl(-{1\over 27} \biggr)=0.
\ee
Let us note that
\be
\omega_c(z)= \omega_2(z) \pm 2 \pi \ri \omega_1(z)- \pi^2,
\ee
where the sign reflects the choice of branch for $\log(z)$. The expressions
(\ref{tc}) and (\ref{omc-def}) define~$t_c(z)$ inside the region of convergence of the series
$\widetilde \varpi_{1,2}(z)$. Outside this region, it is convenient
to use an expression for $t_c$ in terms of a Meijer function
 \be
 t_c (z)= {3 \over 2 \pi} \biggl( \CG(-z) -{\pi^2 \over 6} \biggr),
 \ee
where
 \be
 \CG(z)=\frac{G_{3,3}^{3,2}\left(27 z\left|
\begin{array}{c}
 \frac{1}{3},\frac{2}{3},1 \\
 0,0,0 \\
\end{array}
\right.\right)}{2 \sqrt{3} \pi }-\frac{5 \pi ^2}{18}.
\ee
Another useful expression is~\cite{cesv2,cms}
\begin{gather}
t_c(z) = \frac{2\pi}{3} \biggl( \frac{3\psi}{\Gamma\bigl(\frac{2}{3}\bigr)^3}\, {}_3 F_2 \biggl( \frac{1}{3},\frac{1}{3},\frac{1}{3}; \frac{2}{3},\frac{4}{3}\, \bigg| \psi^3 \biggr) - \frac{9\, \psi^2}{2 \Gamma\bigl(\frac{1}{3}\bigr)^3}\, {}_3 F_2 \biggl( \frac{2}{3},\frac{2}{3},\frac{2}{3}; \frac{4}{3},\frac{5}{3}\, \bigg| \psi^3 \biggr) - 1 \biggr),\!
\end{gather}
where
\be
\psi^3= -{1\over 27z}.
\ee
This flat coordinate has the following power series expansion near the conifold point,
\be
t_c(z)= \frac{1}{\sqrt{3}}\delta + {11 \over 18\sqrt{3}} \delta^2+ {109 \over 243\sqrt{3}} \delta^3+ \cdots.
\ee

\subsection{An integral representation of the domain wall tension}
\label{app-disk-av}

It was pointed out in~\cite{walcher-tadpole} that the domain wall tension for local $\IP^2$
(\ref{TLR}) can be obtained as a~definite integral of the canonical Liouville one-form on the mirror curve. This is the local
version of the general approach of~\cite{mwalcher, walcher-dq, walcher-bcov} to domain wall tensions. Since this is not
fully developed in~\cite{walcher-tadpole}, we provide here some
details for completeness.

The mirror curve of local $\IP^2$ is given by
\be
\label{mirrorc}
\re^x + \re^y + \re^{-y-x} + \kappa=0,
\ee
where $\kappa^3=z^{-1}$ and $z$ is the modulus of local $\IP^2$. It is useful to consider the exponentiated variables
\be
X= \re^x, \qquad Y =\re^y.
\ee
 We note that the equation for the curve can be solved as
 \be
 Y= {X^2 + \kappa X + {\sqrt{ (X^2+ \kappa X)^2- 4 X}} \over 2 X}.
 \ee
 Let us consider the points $p=(X, Y)$ in exponentiated variables given by
\be
p_\pm =\pm \bigl(\kappa^{-1/2}, -\kappa^{-1/2}\bigr),
\ee
which belong to the curve (\ref{mirrorc}). Then, we have that
\be
\label{T-integral}
{1\over \xi} \CT(z)= \int_{X_-}^{X_+} \log(Y) {\rd X \over X}.
\ee
To verify this, we expand in series around $z=0$, to make contact with the expression (\ref{TLR}). The expansion of $y$ reads
\be
\log(Y) =-{1\over 3} \log(z) +\sum_{n \ge 1} z^{n/2} p_n(u),
\ee
where we have introduced the variable $u$ through $X=\kappa^{-1/2} u$, and $p_n(u)$ are Laurent polynomials in $u$. One has, for example,
\be
p_1(u)= u-u^{-1}, \qquad p_2(u)= -\frac{u^2}{2}-\frac{3}{2 u^2}+2.
\ee
One also notes that $p_n(u)$ are even (odd) functions of $u$ for $n$ even (odd). To perform the integral~(\ref{T-integral}),
we have to be careful, since the integrand is singular at $u=0$. We just consider the indefinite integral
\be
\int^u p_n(u') {\rd u' \over u'}= P_n(u),
\ee
and we declare
\be
{1\over \xi} \CT(z)=\int_{-1}^1 \log(Y) {\rd u \over u} =2 \sum_{k\ge 0} z^{k+1/2} P_{2k+1}(1),
\ee
so that only the integrals of odd terms in the series contribute. Then, one verifies that the expansion (\ref{TLR}) is recovered, as claimed in~\cite{walcher-tadpole}.

\subsection*{Acknowledgements}
We would like to thank Johannes Walcher for his insightful comments on a preliminary draft of this paper. We would also
like to thank the anonymous referees for their
helpful comments, which improved very much the quality of the exposition. This work has been
supported in part by the ERC-SyG project ``Recursive and Exact New
Quantum Theory'' (ReNewQuantum), which received funding from the
European Research Council (ERC) under the European Union's Horizon
2020 research and innovation program, grant agreement No. 810573.

\pdfbookmark[1]{References}{ref}
\LastPageEnding

\end{document}